%% file: main.tex
\documentclass[3p]{elsarticle}

\usepackage{lineno,hyperref}
\usepackage{amsmath}
\usepackage{amsfonts}
\usepackage{xcolor}
\usepackage{booktabs}
\usepackage{geometry}
\usepackage{subcaption}
\usepackage{algorithm}
\usepackage{algpseudocode}
\usepackage{adjustbox}
\usepackage{threeparttable}
\usepackage{array}
\usepackage{float}
\usepackage{multirow}

\usepackage{caption}
\usepackage{wrapfig}
\modulolinenumbers[1]
\input{math_commands}

\biboptions{authoryear}

\begin{document}

\begin{frontmatter}

\title{Addressing the alignment problem in transportation policy making: an LLM approach}


\author[mymainaddress]{Xiaoyu Yan}

\author[mymainaddress]{Tianxing Dai}

\author[mymainaddress]{Yu (Marco) Nie\corref{mycorrespondingauthor}}
\cortext[mycorrespondingauthor]{Corresponding author}

\address[mymainaddress]{Department of Civil and Environmental Engineering, Northwestern University, 2145 Sheridan Road, Evanston, IL 60208, USA}

\begin{abstract}

A key challenge in transportation planning is that the collective preferences of heterogeneous travelers often diverge from the policies produced by model-driven decision tools. This misalignment frequently results in implementation delays or failures. Here, we investigate whether large language models (LLMs)—noted for their capabilities in reasoning and simulating human decision-making—can help inform and address this alignment problem.
We develop a multi-agent simulation in which LLMs, acting as agents representing residents from different communities in a city, participate in a referendum on a set of transit policy proposals.  Using chain-of-thought reasoning, LLM agents provide  Ranked-Choice or approval-based preferences, which are aggregated using instant-runoff voting (IRV) to model democratic consensus.
We implement this simulation framework with both GPT-4o and Claude-3.5, and apply it for Chicago and Houston. 
Our findings suggest that LLM agents are capable of approximating plausible collective preferences and responding to local context, while also displaying model-specific behavioral biases and modest divergences from optimization-based benchmarks. These capabilities underscore both promise and limitations of LLMs as tools for solving the alignment problem in transportation decision-making.

\end{abstract}


\end{frontmatter}


\section{Introduction}


Urban transportation policy plays a central role in shaping regional development. Designing effective policy requires access to multidimensional data and a deep understanding of individual preferences across heterogeneous communities. Conventional approaches typically rely on structured mathematical models that identify an optimal policy under specified objectives and constraints. However, these models often rest on rigid assumptions and oversimplified behavioral representations. As a result, they may produce solutions that are analytically tractable yet poorly aligned with public sentiment or the complex realities of policy implementation. This misalignment frequently contributes to delays—or even failures—in policy approval and execution.

Recent advances in large language models (LLMs) offer a promising opportunity to address this alignment problem. Trained on vast corpora of text encompassing news, facts, and human discourse, LLMs possess a rich contextual understanding that could potentially help policymakers infer public preferences and explore trade-offs before implementation. Their ability to interpret unstructured information, reason about competing objectives in natural language, and adapt to specific contexts suggests a new form of decision support that complements the traditional paradigm.

In this study, we implement a multi-agent voting framework to examine the potential of LLMs in supporting transportation policy design. We simulate collective decision-making by deploying autonomous LLM agents as representatives of heterogeneous communities within a large city. These agents participate in a referendum over transit policy proposals involving three levers: a dedicated sales tax for transit services, fare policies, and driver fees (e.g., congestion charges). This design enables us to study how collective preferences form, how trade-offs are negotiated across constituencies, and how democratic mechanisms can be modeled within an AI-driven environment. Crucially, the framework distributes reasoning tasks across agents, enabling scalable deliberation.

To ground the experiment in established transportation planning practices, we incorporate a standard utility-based travel demand model. Travelers, characterized by their daily disposable income, choose between driving and transit, both affected by congestion externalities. The model estimates how each policy scenario impacts travel experience and utility. These outputs are provided to LLM agents to guide deliberation. Additionally, the model yields utility distributions across the income spectrum, allowing us to rank policy alternatives based on normative objectives. These rankings serve as benchmarks to evaluate the LLM agents' choices. We implement both Ranked-Choice and Approval voting to examine how different aggregation rules shape collective outcomes. We also vary the information available to agents to assess how contexts affect decision quality and alignment with model-based benchmarks.

Our research addresses several key questions: (1) To what extent can LLM agents generate coherent and reasonable  policy preferences in a simulated voting environment? (2) How do different voting mechanisms influence  decision outcomes? (3) How  and why do the policies emerged from LLM-based voting  deviate from those recommended by the conventional model? 
(4) How do results generalize across different urban contexts and language models?

We find that LLM-based referendums generally align with model-based benchmarks, selecting policies that reflect similar priorities despite differences in detail. A notable exception is that LLM agents consistently display stronger aversion to taxation than the model-prescribed optima. Our results also indicate that voting behavior varies by model: GPT-4o produces more consistent and decisive patterns, whereas Claude-3.5 yields more nuanced responses. Sentiment analysis helps explain this contrast, showing that GPT-4o agents expressed uniformly positive tones in their rationales, while Claude-3.5 agents conveyed a wider range of sentiment. Nonetheless, both converge on similar average preferences, and the top-ranked policies remain stable across voting methods. Regression analysis further confirms that LLM agents—regardless of the model—express preferences largely consistent with intuitive expectations, offering reassuring evidence of robustness. Finally, we observe context sensitivity: in Houston, GPT-4o agents favor policies with lower tax burdens and higher driver fees than in Chicago, hinting at an implicit awareness of local sociopolitical conditions.


The remainder of this article is organized as follows: Section 2 reviews related studies; Section 3 describes the proposed methodology; Section 4 outlines the experiment setup, presents and interprets simulation, sentiment analysis and regression results; and Section 5 summarizes the finding and comments on future research directions.

\section{Related studies}

\subsection{Transportation Planning}

Travel forecasting models have served as the analytical backbone of urban transportation planning for over half a century \citep{beckmann1956studies,manheim1979fundamentals,weiner1997urban,meyer2001urban, boyce2015forecasting}. The dominant paradigm—originating from the four-step modeling framework developed during the postwar era—was designed to forecast long-term travel demand under assumptions of population growth, land use change, and infrastructure expansion.

Since the 1970's, travel forecasting models have been grounded on a micro-economic foundation \citep{mcfadden1973conditional,ben1985discrete}. Rooted in random utility theory, they typically assume that travelers make travel choices (destination, schedule, mode, and route) to maximize a latent utility function composed of quantifiable attributes such as monetary cost, travel time, and socio-demographic attributes. While convenient for statistical estimation and instrumental in shaping regional investments, this approach has faced mounting criticism over its lack of falsifiability, excessive complexity, and inability to produce accurate or socially aligned forecasts \citep{nie2025brief}.

A fundamental limitation of conventional models is their behavioral rigidity. Travelers are modeled as utility maximizers with stable preferences, even though in practice, people make decisions under uncertainty, social influence, habit, and bounded rationality \citep{arrow1966exposition,kahneman1979prospect,simon1990bounded}. They are also poorly equipped to anticipate structural shifts in travel patterns \citep{polak_comment_1987,hartgen_hubris_2013}. Because they are calibrated on historical data and assume parameter stability, they tend to miss the impact of exogenous shocks and long-term societal transformations.

The mathematical tractability of models often comes at the expense of public alignment. Policy-making or analysis tools built on travel forecasting models are generally designed to optimize an objective function—usually efficiency or welfare—subject to resource constraints. But they do little to capture the complexity of public values, normative trade-offs, or democratic deliberation. As a result, model-generated “optimal” solutions can be misaligned with public preferences, leading to friction during implementation. This issue has been described as the \textit{technocratic disconnect} \citep{forester1989planning}, wherein planners rely on technical outputs that lack legitimacy in the public sphere. This disconnect stems partly from the dominance of a planning process purported to be value-neutral and science-driven \citep{Wachs_1982,wachs1989when}. In reality, however, planning and public policymaking are inevitably political \citep{rittel1973dilemmas}, as they involve value judgment, contested narratives, strategic framing, and the exercise of power—not just technical problem-solving.

For example, a new highway project that maximizes regional travel time savings may disproportionately benefit affluent suburban commuters while imposing burdens on transit-dependent urban residents. Indeed, traditional travel forecasting models are limited in their ability to reflect such distributive concerns, as they are often driven implicitly by simplistic normative values such as utilitarianism \citep{martens2017transport,dai2024fare}.

To these fundamental challenges we must add the inconvenient fact that travel forecasts often suffer from significant errors, especially in high-profile infrastructure projects. As early as the 1980s, \cite{williams_travel_1982} described the pursuit of forecasting accuracy in travel behavior as a “delusion” that many in the field had already given up. \cite{flyvbjerg2003megaprojects} found that approximately 85\% of rail projects and half of all road projects had demand forecasts that deviated by more than 20\% from actual outcomes. His message is blunt and devastating: “don’t trust traffic forecasts, especially for rail.” Echoing that dire warning, \cite{hartgen_hubris_2013} estimated that the likely error margin for 20-year demand forecasts on major road projects is at least ±30\%, with some estimates reaching ±40–50\% even over shorter time horizons.

The deficiencies identified above underscore the growing need for complementary tools that can more tightly incorporate public sentiment and democratic deliberation into the decision-making process. Our study is motivated by the hope that LLMs might be positioned to play this role. We next turn to the recent applications of LLMs in related contexts, including in transportation planning.

\subsection{Applications of LLMs in simulating complex human decision-making processes}

Social scientists have increasingly turned to LLMs to simulate human decision-making in realistic social contexts. Unlike conventional models that rely on explicit structural assumptions, LLMs leverage vast amounts of contextual knowledge acquired through pretraining to  infer relationships and evaluate trade-offs. For example, \citet{han2024mining} showed that LLMs respond adaptively to contextual prompts in classical decision-making tasks, implicitly identifying relevant factors and exhibiting a degree of behavioral flexibility beyond that of traditional structural models. Similarly, \citet{ross2024llm} proposed a utility-theoretic framework for mapping behavioral biases in LLMs across canonical economic games. They found that LLM decision patterns diverge from both rational “homo economicus” and human baselines, while remaining sensitive to prompt design.

A growing body of work has explored the potential of LLMs to reproduce realistic cognitive processes and social behaviors. \citet{park2023generative} demonstrated that persona-driven LLM agents can emulate social influence and opinion dynamics, providing a platform to study emergent collective behaviors. In a sequel, \citet{park2024generative} assessed the alignment between LLM-generated survey responses and empirical human data, finding strong correlations that support their use as proxies in behavioral experiments. These findings suggest that LLM-based simulations can serve as ``what-if” laboratories to test public responses to hypothetical policies, communications, and environmental shocks.

Beyond theoretical experimentation, LLMs are also being used to inform practical decision-making. Researchers have deployed LLM agents to model strategic behavior in auction design \citep{chen2023put}, simulate legal reasoning in court-like settings \citep{chen2024agentcourt}, and develop macroeconomic forecasting tools that generate scenario-based projections \citep{li2023econagent}. 

\subsection{LLMs in urban and transportation planning}

A few studies have explored the potential of LLMs in decision-making support for urban planning. For example, \citet{zhou2022large} showed that LLMs can match—and in some cases exceed—the performance of traditional reinforcement learning methods when applied to complex city planning tasks. 
In a subsequent research, \citet{zhou2024large} proposed a multi-agent framework in which LLM-powered agents simulated community residents with diverse daily needs. Using a structured fishbowl discussion mechanism, these agents deliberated collectively to allocate land in a way that balanced competing priorities. The experiments highlighted the capacity of LLMs to facilitate inclusive, participatory planning processes in contexts characterized by value pluralism and trade-offs.

Along similar lines, \citet{ni2024planning} developed a dynamic, closed-loop planning system in which simulated resident agents provided real-time evaluations of planning proposals, while a planner agent iteratively adjusted its strategies in response. This architecture modeled the adaptive and participatory character of human-centered planning and demonstrated how LLMs could be used to approximate stakeholders' feedback and engagement in complex urban systems.

Taken together, these studies suggest that LLMs offer a distinctive advantage in urban planning by integrating heterogeneous perspectives from simulated individuals and representative groups, capturing the value pluralism inherent in real communities, and adapting decisions dynamically through iterative simulations. In doing so, LLM-based frameworks not only facilitate more inclusive and participatory planning processes but also enhance the adaptability and responsiveness of decision-making in complex urban environments.

\subsection{Research gap}

While LLMs have shown promise in simulating human behavior and supporting participatory planning, existing studies have not examined how their collective decisions compare to outcomes produced by conventional forecasting tools—especially in transportation planning. 
To the best of our knowledge, this alignment problem—how well LLM-generated decisions reflect public preferences versus traditional models—has not been systematically studied. Our work fills this gap by directly comparing the policy preferences expressed by LLM agents with the outputs of a utility-based travel demand model.

Methodologically, we also depart from prior studies by framing the policy decision as a referendum. Instead of relying on a single agent or planner, we simulate a vote among heterogeneous agents representing different communities. This structure better captures how real-world decisions are made: not by optimization, but by aggregating diverse and sometimes conflicting interests.

\section{Methodology} \label{sec:methodology}
We consider a stylized city composed of $\mathcal{I}=\{1,\cdots,I\}$ communities. The city's transportation agency  is evaluating a policy change to its current transit system, which consists of three \textit{levers}: a flat fare paid by each rider (denote by $r$), a dedicated sales tax levied on all residents ($t$), and a per-trip fee paid by drivers ($\tau$).  Accordingly, use a tuple $\vp_k = \{r_k, t_k, \tau_k\}$ to denote the values of a policy $k\in \mathcal{K}=\{1,\cdots,K\}$.
To streamline the decision process, the agency defines three levels—low ($l$), medium ($m$), and high($h$)—for each of the three policy levers. In other words, $r_k, t_k$, and $\tau_k$ must take one of three values contained in  set $\{l, m, h\}$. This results in a policy set $\mathcal{K}$ consisting of $K=27$ distinct policy proposals.

The analysis proceeds in two stages. First, the agency evaluates the $K$ policies using a conventional transit policy design model. This model provides estimates of performance metrics such as travel times, trip costs, and congestion levels. Second, the same set of policies is submitted to a multi-agent simulation framework powered by a large language model (LLM). This simulator emulates a city-wide referendum, where agents representing different communities deliberate and vote, informed by, among other things, the output of the conventional model.

The next three sections describe the components of our approach. Section \ref{sec:ConventionalModel} presents the transit design model. Section \ref{sec:LLMExperiment} describes the LLM-based simulation framework. Section \ref{sec:Evaluation} introduce several metrics used to compare outcomes across the two approaches.

\subsection{Transit design model} \label{sec:ConventionalModel}

We take the transit design model from \citet{dai2024fare}, which is built atop of a square city with a grid street network and evenly distributed residents and travel demand. Residents of the city differ by income level, which shapes their travel choices and how they experience transportation costs and benefits.

A bus network is overlaid on the city’s street grid. Its service quality is governed by three operational parameters: headway (how frequently buses arrive), stop spacing (distance between adjacent stops), and route spacing (distance between adjacent bus lines). These design elements collectively determine the level of service of the bus system. Buses and private cars share the same roads, and there is no dedicated transportation infrastructure for transit.

The bus system is financed through a combination of rider fare $r$, a local sales tax on residents $t$, a fee paid by drivers $\tau$ (e.g., a congestion charge or fuel tax), and an exogenous government subsidy. The transit agency must determine both the service configuration and the financing strategy to balance service quality with fiscal feasibility.

Each resident chooses between two travel modes—bus or car—for a fixed number of daily trips. This choice depends on both cost and accessibility. Accessibility, in turn, is shaped by the level of service of the bus system (for bus riders) and by network travel speeds (for both bus riders and drivers). Travel speeds themselves are influenced by congestion, which is in turn affected by mode choice.

Congestion is modeled using a macroscopic fundamental diagram \citep{geroliminis2008existence} that links traffic density to average speed. As more residents choose to drive, network congestion increases, reducing travel speeds for all travelers. Given their smaller share of total traffic, buses are assumed not to contribute significantly to congestion. Moreover, the system is evaluated under peak travel conditions, when the performance is most critical.

\begin{figure}[h]
    \centering
    \includegraphics[width=0.75\linewidth]{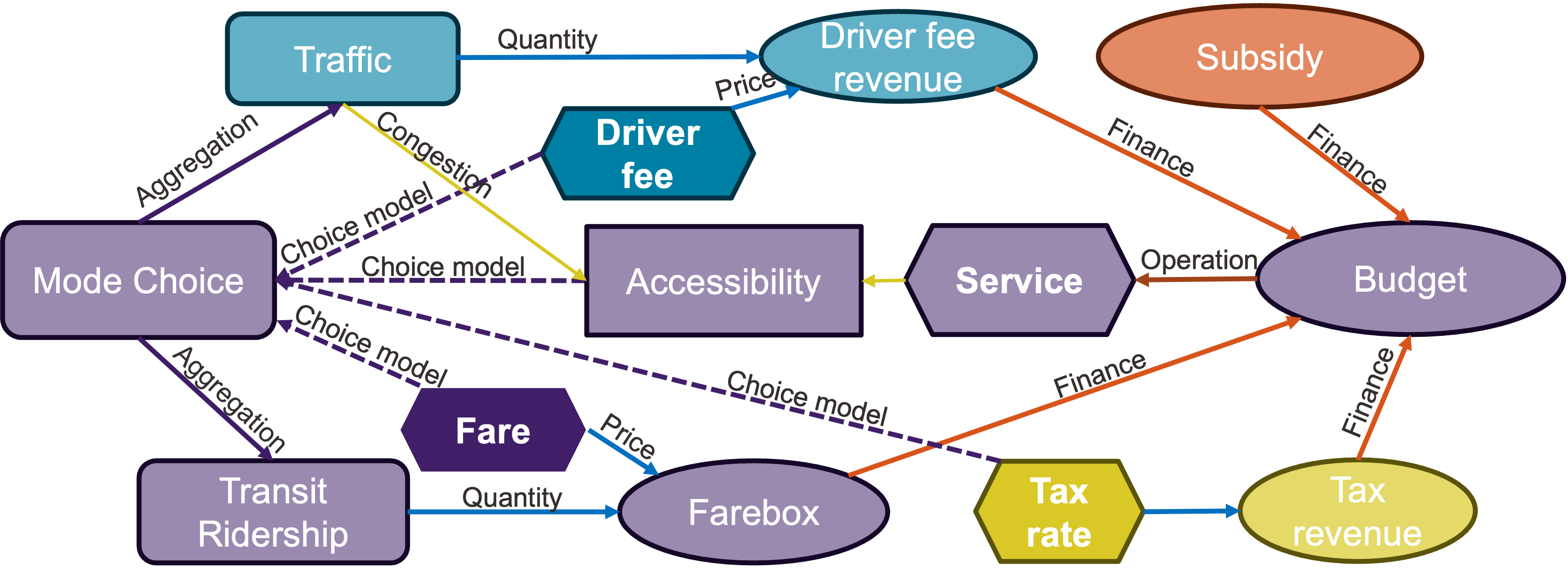}
    \caption{Joint design of public transit service and policy.}
    \label{fig:transit_frame}
\end{figure}

Figure \ref{fig:transit_frame} illustrates the structure of this integrated design model. Key decision variables include both service parameters (headway, stop spacing, and line spacing) and policy instruments (fares, driver fees, and tax rates). At the core of the model is mode choice, which depends on income, cost, and accessibility. These individual choices aggregate into system-level travel patterns that shape congestion, influence accessibility and affect overall revenue for transit through a feedback loop. The agency then adjusts its service and financial decisions, subject to a budget constraint requiring that operating costs be covered by the sum of fares, taxes, driver fees, and subsidies.

The above model is employed to evaluate system objectives, transit ridership, and distributive effects after a transit policy $k\in\mathcal{K}$ is implemented. For each policy $k\in \mathcal{K}$, the corresponding transit mode share is denoted as $\gamma_k$.  We compute two normative objectives: the total utility of all travelers $U_k$ (a utilitarian objective) and the utility of the most disadvantaged traveler $u_k$ (an egalitarian objective).  Roughly speaking, $U_k$ measures efficiency whereas $u_k$ gauges distributive effects. Another metric, a more direct indicator of distribution effects, is the Gini index, denoted as $G_k$.  These  metrics serve to evaluate the performance of our LLM-based simulation framework.

\subsection{LLM-based multi-agent simulation}
\label{sec:LLMExperiment}
\begin{figure}[ht!]
    \centering
    \includegraphics[width=0.75\linewidth]{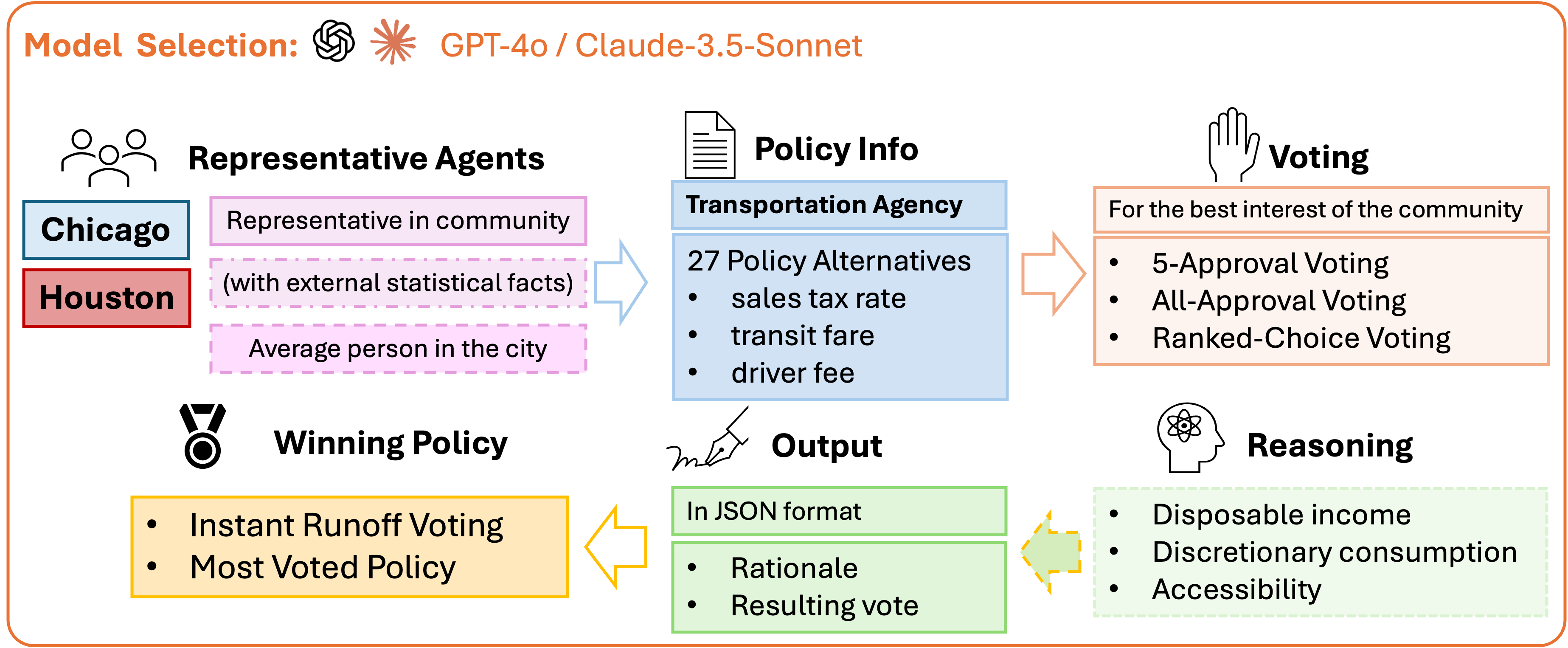}
    \caption{Framework of the multi-agent LLM simulation}
    \label{fig:framework}
\end{figure}

We implement a multi-agent simulation framework in which LLM agents serve as representatives of communities within a city, as illustrated in Figure \ref{fig:framework}. Each agent is tasked with evaluating and voting on the $K$ transit policies, as described at the beginning of Section \ref{sec:methodology}. Recall that each policy $k\in \mathcal{K}$ is formed by choosing the three policy levers, a transit fare $r_k$, sales tax rate $t_k$, and driver fee $\tau_k$, $ { \{t, r, \tau\}}\in \mathcal{L}$ , from one of three levels $ q \in\{l, m, h\}$ . 

Two state-of-the-art LLMs are used in our simulation: GPT-4o-2024-08-06 from OpenAI \citep{openai2025chatgpt} and Claude-3.5-Sonnet-20241022 from Anthropic \citep{anthropic2024claude35sonnet}. Both models are used with temperature set to zero to ensure reproducibility and minimize stochastic variation in the outputs.

We construct three types of agents to examine the role of contextual information in shaping LLM decision-making:
\begin{enumerate}
    \item[1.] \textbf{Community-based agents} represent individual communities and acted based solely on contextual information acquired in pretraining.
    \item[2.]  \textbf{Knowledge-augmented agents} build on the first configuration, but are additionally prompted using localized demographic and economic data—such as average household expenditures, transit reliance, and income levels—to test whether broader contextual grounding improves decision-making.
    \item[3.]  \textbf{City-average agents} simulate a generic “average” resident of the city, without community-specific setting and contexts.
\end{enumerate}

Each agent $i \in \mathcal{I}$ is instructed to evaluate the $K$ policies and cast votes based on what best serves its community’s interests. Let $\mathbf{v}_i = \{v_{i,0}, \dots, v_{i,K-1}\}$ denote the vote cast by agent $i$. Since there is no universally accepted voting rule for simulated multi-agent environments, we experiment with three collective decision rules commonly studied in  literature \citep{yang2024llm}:

\begin{itemize}
\item[1.] \textbf{5-Approval voting}: Each agent approves exactly five proposals. Formally, $v_{i,k} \in \{0,1\}$ for all $k$, with $\sum_{k=0}^{K-1} v_{i,k} = 5$ for each $i \in \mathcal{I}$.
\item[2.] \textbf{All-Approval voting}: Agents may approve any number of proposals they deem acceptable, i.e., $v_{i,k} \in \{0,1\}$, without any constraint on the number of approvals.
\item[3.] \textbf{Ranked-Choice voting}: Each agent assigns a unique rank to up to five proposals. Formally, \( v_{i,k} \in \{1, 2, 3, 4, 5, \text{null}\} \), where lower values indicate higher preference, and each agent assigns at most one value from \( \{1,2,3,4,5\} \) to any proposal. That is, for all \( i \in \mathcal{I} \), \( |\{k : v_{i,k} \in \{1,2,3,4,5\}\}| \leq 5 \) and \( v_{i,k} \neq v_{i,k'} \) for \( k \neq k' \) whenever both are ranked.
\end{itemize}

To simulate thoughtful decision-making, we adopt a structured chain-of-thought prompting strategy. Before casting a vote, each agent is explicitly instructed to reason through three key considerations:
\begin{itemize}
    \item[1.]         \textbf{Disposable income}: How the proposed policy affects residents’ income after taxes, fares, and fees.
    \item[2.]      \textbf{Discretionary consumption}: How much income is left for non-essential goods and services.
    \item[3.]      \textbf{Accessibility}: The ability to reach daily destinations—such as work, shopping, or healthcare—under the transit and congestion conditions implied by each policy.
\end{itemize}
These instructions are meant to elicit deliberative reasoning akin to what a civic-minded resident might engage in when voting in a real referendum. Rather than prescribing fixed formulas, we allow the LLMs to internalize trade-offs and form judgments by drawing on contextual knowledge acquired through pretraining. Each agent produces a structured output in JSON format that includes its community ID, a written rationale for its decision, and the resulting vote (either ranked or approved alternatives).

To ensure consistency and elicit transparent and structured reasoning from each LLM agent, we carefully crafted both system and user prompts, provided in Appendix~\ref{fig:prompt_exp}. The system prompt frames the agent’s role as a representative of a specific community area, instructing it to think step by step and submit a ranked list of preferred policy proposals, justified by the three factors mentioned earlier.  The user prompt introduces the policy referendum context, details the 27 candidate policy options, and provides key performance metrics including cost and travel time by mode. It also highlights how different policy levers impact both travel behavior and household budgets.

\subsection{Evaluation metrics}
\label{sec:Evaluation}
Several evaluation metrics are used extensively in the simulation experiments. Thus, we provide an introduction here for the convenience of the reader. 

\subsubsection{Winner of referendum and policy means of selected policies}
In the simulation, voting is carried out in multiple rounds to ensure the results are stable.  Accordingly, we let $\vv_i^j = \{v_{i,k}^j\}$ be the vote of agent $i\in \mathcal{I}$ in round $j$.  In each round, we can determine the winner and compute the mean values of the polices voted for. The policy means can then be
averaged over the rounds.
\begin{itemize}
\item[1.] \textbf{5-Approval voting}: The winner in each round is the policy that receives the most votes, i.e., $k^* = \mathrm{argmax}_k \sum_{i\in\mathcal{I}}v_{i,k}$. We can compute the mean values of all approved polices as 

\begin{align}
\bar{\vp}^j_{A} = 
\frac{\sum_{k \in \mathcal{K}} \sum_{i \in \mathcal{I}} \vp_{k} \, v_{i, k}^{j}}
     {\sum_{k \in \mathcal{K}} \sum_{i \in \mathcal{I}} v_{i, k}^{j}}.
\end{align}
\item[2.] \textbf{All-Approval voting}: Both the determination of the winner and the calculation of the mean policy values are identical to 5-Approval voting. 
\item[3.] \textbf{Ranked-Choice voting}: In Ranked-Choice voting, the winner is determined using the instant-runoff procedure, which proceeds as follows:

\begin{itemize}
    \item[a.]     First-choice tally: Each agent's top-ranked proposal (i.e., the one with $v_{i,k}=1$) receives one vote.
    \item[b.]     Majority check: If any proposal receives a majority of the first-choice votes, it is declared the winner.
    \item[c.]     Elimination: If no proposal has a majority, the proposal with the fewest first-choice votes is eliminated.
    \item[d.]     Reallocation: For agents whose top choice was eliminated, their vote is transferred to their next-highest-ranked remaining proposal (e.g., the one with $v_{i,k}=2$, if still in contention).
    \item[e.] Repeat: Steps b–d are repeated until one proposal obtains a majority of the active votes.
\end{itemize}
Assuming all agents cast complete ranked ballots of five proposals in referendum round $j$, let $\vp_{i,s}^j$ denote the value vector associated with the policy ranked at position $s \in \{1,2,3,4,5\}$ by agent $i \in \mathcal{I}$. The mean policy value at rank $s$ in round $j$ is then defined as:
\begin{align}
\bar{\vp}_s^j = \frac{1}{|\mathcal{I}|} \sum_{i \in \mathcal{I}} \vp_{i,s}^j.
\end{align}
This yields a ranked list of average policy vectors $\{\bar{\vp}_1^j, \bar{\vp}_2^j, \ldots, \bar{\vp}_5^j\}$, which summarizes aggregate preferences across agents at each rank level.

\end{itemize}

\subsubsection{Entropy}

We use entropy to quantify the variation in voting outcomes across agents in round $j$, where the voting profile is denoted by 
$\vv^j=\{v_i^j\}$. For this purpose, let $P_k^j$ represent the empirical probability of policy 
$k$ being selected in round $j$. This probability is obtained by dividing the number of times policy 
$k$ is voted for (regardless of ranking position, if applicable) by the total number of votes cast in that round. Note that each community agent may contribute multiple votes.
For 5-Approval Voting or All-Approval Voting


\begin{align}
P_k^j = \frac{\sum_{i \in \mathcal{I}} v_{i,k}^j}{\sum_{k \in \mathcal{K}}\sum_{i \in \mathcal{I}} v_{i,k}^j} .
\end{align}

For Ranked-Choice Voting, let $\mathbb{I}(\cdot)$ be the indicator function. We have


\begin{align}
P_k^j = \frac{1}{|\mathcal{I}|} \sum_{i \in \mathcal{I}} \mathbb{I}\left( \text{policy } k(i,j,s), s\in \{1,2,3,4,5\} \right),
\end{align}
where $k(i,j,s)$ is the index of the policy ranked at position $s$ by agent $i$ in round $j$.
The policy entropy in round $j$ is given by
\begin{align}
E^j = - \sum_k P_k^j \log_2 P_k^j.
\end{align}
\noindent
The average of the policy entropy among $J$ rounds is defined as $\bar{E}$.

We further define entropy for each of the three policy levers in a policy $k$, which can only take one of the three values included in the set $\{l,m,h\}$. Let $P_{x,y}^j$ be the probability of choosing policy lever $x\in\{r, t, \tau\}$ at level $y\in\{l,m,h\}$ in round $j$, which can be computed by noting that the policy set $\mathcal{K}$ corresponds to a tensor product of the lever set and the level set.  Accordingly, we can define the entropy for each policy lever $x$ in round $j$ as
\begin{align}
e_{x}^j =  - \sum_{y\in \{l,m,h\}} P_{x,y}^j \log_2 P_{x,y}^j, \forall x.
\end{align}
For the Ranked-Choice voting, to further see the variations within votes of same rank, we define the entropy for each lever $x$ in round $j$, conditioned on the vote rank $s$, as 
\begin{align}
e_{x|s}^j =  - \sum_{y\in \{l,m,h\}} P_{x,y|s}^j \log_2 P_{x,y|s}^j, \forall x,
\end{align}
where $P_{x,y|s}^j$ is the probability of choosing policy lever $x\in\{r, t, \tau\}$ at level $y\in\{l,m,h\}$ in round $j$ among all the rank $s$ votes. Finally, we use $\bar{e}_{x|s}$ to denote the average entropy of lever $x$ over all $J$ rounds.



\subsubsection{VADER sentiment analysis}
We use VADER (Valence Aware Dictionary and sEntiment Reasoner) to quantify the sentiment expressed in the rationale texts generated by LLM agents \citep{hutto2014vader}. VADER is a lexicon-based sentiment analysis tool that applies heuristic rules to adjust for features such as punctuation, capitalization, negation, and intensity modifiers. For a given rationale text $T_i$ from agent $i$, VADER outputs a compound sentiment score $S_i \in [-1, 1]$, defined as:
\begin{align}
S_i = \frac{o_i - n_i}{\sqrt{(o_i - n_i)^2 + \alpha}} = \text{VADER}(T_i),
\end{align}
where $o_i$ and $n_i$ represent the total positive and negative polarity scores assigned by the VADER lexicon, and $\alpha = 15$ is a normalization constant. These polarity scores are computed by summing the sentiment intensities of individual words in $T_i$, as matched against VADER’s predefined dictionary, which maps words and phrases to sentiment values.

For each agent $i\in \mathcal{I}$, we compute $S_i$ and analyze the distribution across all agents to gain insight into the emotional tone and preference strength expressed during deliberation. This analysis helps reveal alignment trends, sentiment consistency, and the degree of conviction in agents’ reasoning about transit policy proposals.

\section{Results}
In what follows, we present our results in four parts. Section~\ref{sec:res_model} begins with a benchmark analysis based on the conventional model. Section~\ref{sec:res_chicago} presents the LLM-based simulation results for Chicago, analyzing the emergent patterns of agent preferences, reasoning, and voting outcomes, followed by a comparison in Section \ref{sec:gpt_vs_claude} of the two LLMs (GPT-4o and Claude-3.5)  based on sentiment analysis.  In Section \ref{sec:res-preferences}, we conduct a regression analysis to better understand LLM agents' preferences, as revealed in their voting behaviors.   Finally, Section~\ref{sec:res_houston} assesses the generalizability of the framework by comparing simulation results between Chicago and Houston. 

All experiments were conducted on a MacBook Air equipped with an Apple M2 chip and 10 GB of unified memory, running macOS. The simulation framework was implemented and executed in Python (version 3.11).

\subsection{Model-based results}
\label{sec:res_model}
As discussed in Section~\ref{sec:ConventionalModel}, a core component of the transit design model developed by \citet{dai2024fare} is utility-based mode choice, in which residents select between transit and driving to maximize individual utility. Mode-specific utility depends on a range of factors, including income, population density, traffic congestion, transit service levels and fares, as well as policy instruments such as dedicated taxes and driver fees. The model is calibrated by aligning key statistics  with empirical data. For the Chicago case study, \citet{dai2024fare} target four statistics: (i) transit mode share, (ii) daily ridership, (iii) farebox recovery ratio (i.e., the percentage of transit operating budget covered by fare revenue), and (iv) total transit budget.

For consistency, we adopt the calibrated parameter set from \citet{dai2024fare} for the Chicago simulations. For the Houston case study, however, the model must be re-calibrated to match city-specific conditions. Table~\ref{tab:calibration} summarizes the calibration results for both cities, with data sources noted in parentheses. Across all four calibration statistics, the model achieves a close match with empirical values. Furthermore, while the congestion index (defined as the ratio of congested speed to free-flow speed) was not explicitly targeted during calibration, its predicted value also aligns well with observed data—offering additional support for the validity of the calibrated model.
 \begin{table}[!ht]
    \centering
    \small
    \caption{Key statistics produced by the calibrated model vs. empirical data.} \label{tab:calibration}
    \begin{tabular}{l l l l l}
    \hline
        & \multicolumn{2}{c}{Chicago} & \multicolumn{2}{c}{Houston}\\ 
        & Data & Model & Data & Model \\ \hline
        Transit mode share & 30\% \citep{comeaux2021pre} & 29\% & 4\% \citep{CensusR2023} & 4\% \\ 
        Daily ridership & 1.47 million \citep{NTD2020} & 1.46 million & 0.29 million \citep{NTD2020} & 0.30 million\\ 
        Farebox & 43\% \citep{NTD2020} & 44\% & 12\% \citep{NTD2020} & 11\%\\ 
        Peak hour budget & \$441/$\text{km}^2$ \citep{NTD2020}& \$441/$\text{km}^2$ & \$125/$\text{km}^2$ \citep{NTD2020}& \$125/$\text{km}^2$\\ 
        Congestion index  & 0.77 \citep{FHWA}& 0.74 & 0.79 \citep{FHWA} & 0.77 \\ \hline
    \end{tabular}
\end{table}

In all experiments, the policy space is defined by three levers, each with three discrete levels:

\begin{itemize}
    \item[1.] \textbf{Fare:} \( l \) – \$0.75/trip; \( m \) – \$1.25/trip; \( h \) – \$1.75/trip;
    \item[2.] \textbf{Sales tax:} \( l \) – 0.5\%; \( m \) – 1\%; \( h \) – 1.5\%;
    \item[3.] \textbf{Driver fee:} \( l \) – \$0/trip; \( m \) – \$0.50/trip; \( h \) – \$1/trip.
\end{itemize}

Table~\ref{tab:policy_table} in Appendix lists the resulting 27 policy combinations and their associated performance metrics as computed by the calibrated model for the Chicago case. A subset of these metrics—specifically, mode-specific travel times and trip costs—are provided to the LLM agents during simulation to guide their voting behavior. Other model outputs, such as average utility (\( U_k \)), minimum utility (\( u_k \)), the Gini index (\( G_k \)), and transit mode share (\( \gamma_k \)), are used strictly for evaluation purposes. Agents have no access to these metrics, nor to any internal structure of the model, including the utility function or the welfare computation process.

Among these metrics, \( U_k \) serves as a proxy for system efficiency, while \( u_k \), \( G_k \), and \( \gamma_k \) reflect various aspects of distributional equity. For completeness, Table~\ref{tab:houston_policy} provides the same information for the Houston case study.


\begin{figure}[ht!]
    \centering
    \begin{subfigure}[b]{0.32\textwidth}
        \centering
        \includegraphics[width=\textwidth]{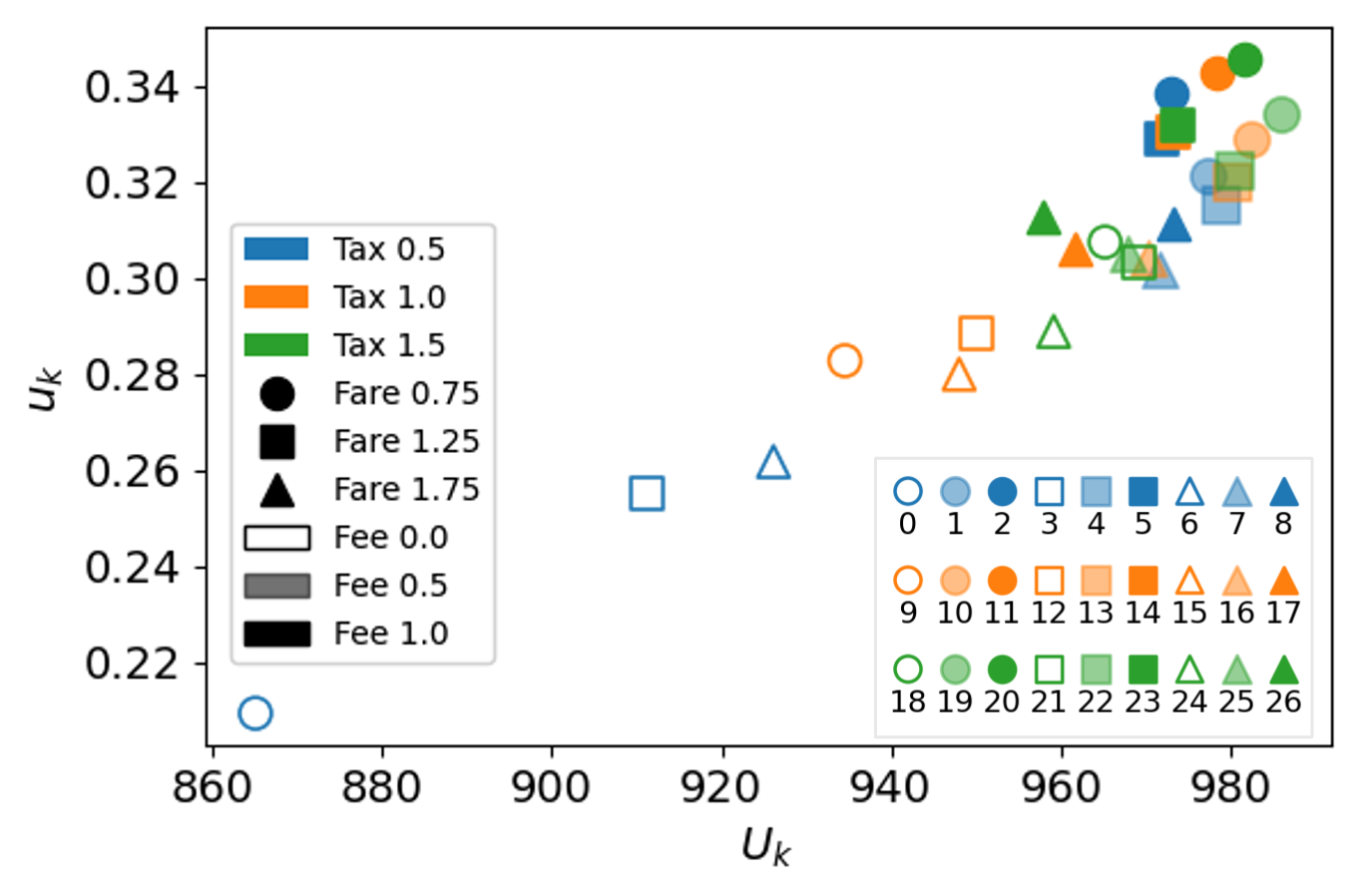}
        \vspace{-10mm}
        \label{fig:umin}
    \end{subfigure}
    \begin{subfigure}[b]{0.32\textwidth}
        \centering
        \includegraphics[width=\textwidth]{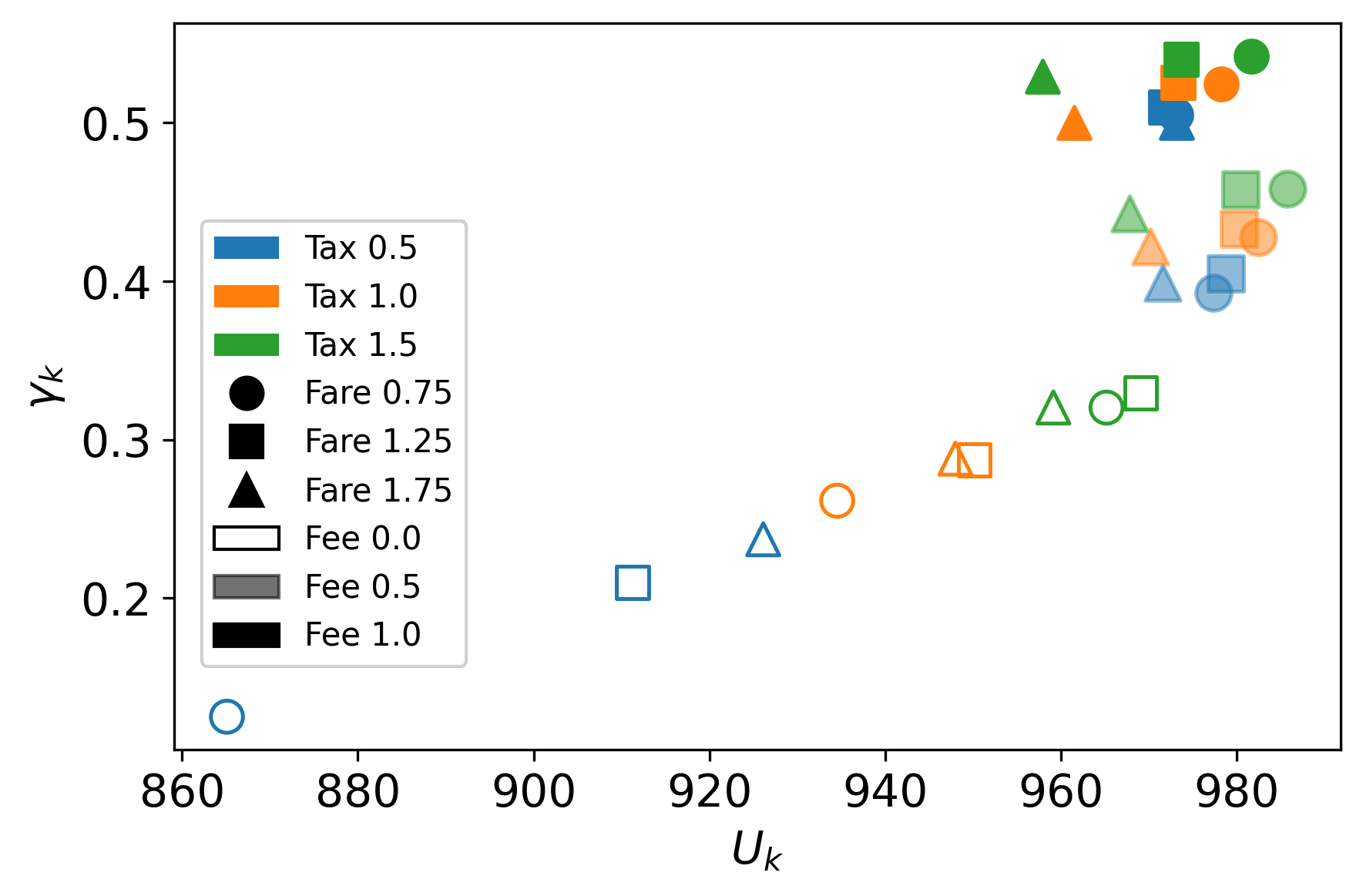}
        \vspace{-10mm}
        \label{fig:transit}
    \end{subfigure}
    \begin{subfigure}[b]{0.32\textwidth}
        \centering
        \includegraphics[width=\textwidth]{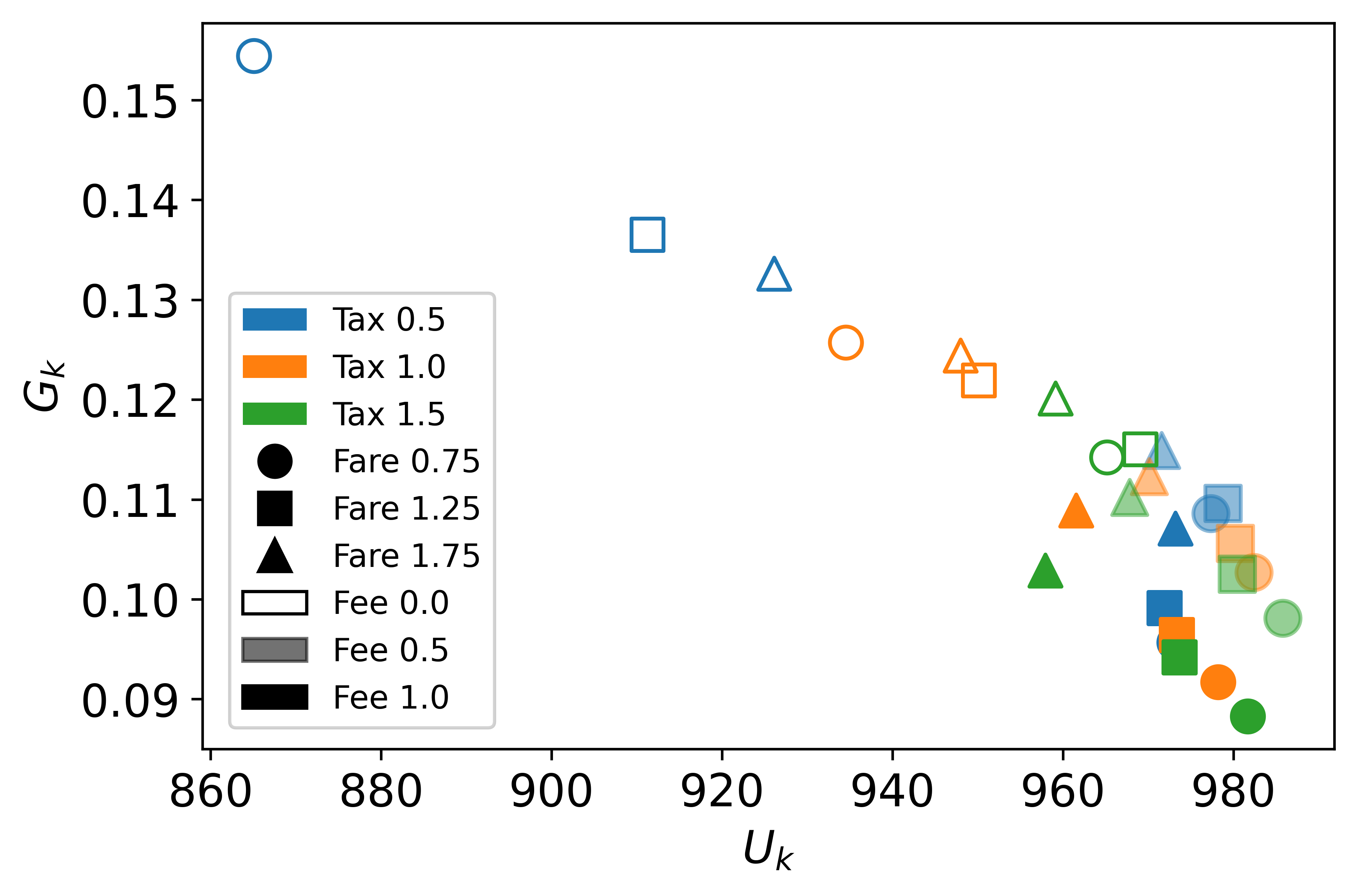}
        \vspace{-10mm}
        \label{fig:gini}
    \end{subfigure}
    \caption{Comparison of $u_k$, $\gamma_k$, and $G_k$ against $U_k$ for different policy configurations in Chicago.}
    \label{fig:util_comparison_chicago}
\end{figure}

The status quo (\textit{Policy 12}, marked by the orange hollow square) does not lie on the Pareto frontier, which is defined by the upper-right envelope in the left and middle plots and the lower-right envelope in the right plot of Figure~\ref{fig:util_comparison_chicago}. By increasing the sales tax and driver fee while keeping the transit fare constant or reduced, one can simultaneously increase \( U_k \) and \( \gamma_k \), and decrease \( G_k \), thereby achieving both more utilitarian and more egalitarian outcomes. The underlying mechanism is intuitive: greater transit subsidy via taxes and driver fees enhances the level of service, induces a shift from driving to transit, and ultimately reduces congestion—improving the overall efficiency of the transportation system.

In terms of model-optimal outcomes, Figure~\ref{fig:util_comparison_chicago} shows that the \textit{Utilitarian} solution corresponds to \textit{Policy 19} (green transparent circle: 1.5\% tax rate, \$0.75 transit fare, \$0.50 driver fee), which maximizes the total utility \( U_k \). However, the minimum utility \( u_{19} \) under this policy is not the highest among all \( u_k \), indicating that maximizing \( U_k \) does not necessarily benefit the most disadvantaged travelers. By contrast, the \textit{Egalitarian} solution is \textit{Policy 20} (green solid circle: 1.5\% tax rate, \$0.75 transit fare, \$1.00 driver fee), which achieves the highest minimum utility, the greatest transit mode share, and the lowest Gini index among all policy options.

In general, lowering the transit fare tends to move outcomes closer to the Pareto frontier. Conversely, when the sales tax is low and no driver fee is levied—as in \textit{Policy 0} (blue hollow circle)—the resulting transit budget is insufficient to sustain a reasonable level of service. Nonetheless, we observe a tight cluster of solid and transparent points located near the Pareto frontiers, suggesting that many policies produce outcomes similar to the optimal ones defined by utility-based objectives. If one were to adopt an alternative social welfare function, any of these clustered policies could emerge as optimal. This diversity sets the stage for the LLM-based voting experiment that follows.


\subsection{LLM referendum for Chicago}
\label{sec:res_chicago}

Chicago has 77 community areas. Thus, the simulated referendum is participated by 77 communities members represented by LLM agents. We begin by comparing the outcomes produced by the three voting methods described earlier. 
We then explore how contextual information and knowledge embedded in the prompts influence agents' voting patterns. 
All results reported in this section are generated using the GPT-4o model. The comparison with Claude-3.5 will be discussed later. 
 

\subsubsection{Results from different voting methods}

For each of the three voting types --- Ranked-Choice voting, 5-Approval policy and All-Approval policy --- we keep the main prompt structure the same, and change the instruction for the voting rule as follows: 
\begin{enumerate}
    \item Ranked-Choice voting:  You are allowed to choose five policies and submit them as a ranked list. 
    \item 5-Approval: You are allowed to choose five proposals and submit them as a list. 
    \item All-Approval: You should vote all the policies that you agree and submit them as a list. 
\end{enumerate}
The agents vote independently, and when all votes are cast,   the winner is determined using the methods described earlier. For a sample response obtained by each voting method, the reader is referred to  ~\ref{sec:sample_response_all_approve} - ~\ref{sec:sample_response_rk}.

\begin{figure}[ht!]
    \centering
    \includegraphics[width=0.75\linewidth]{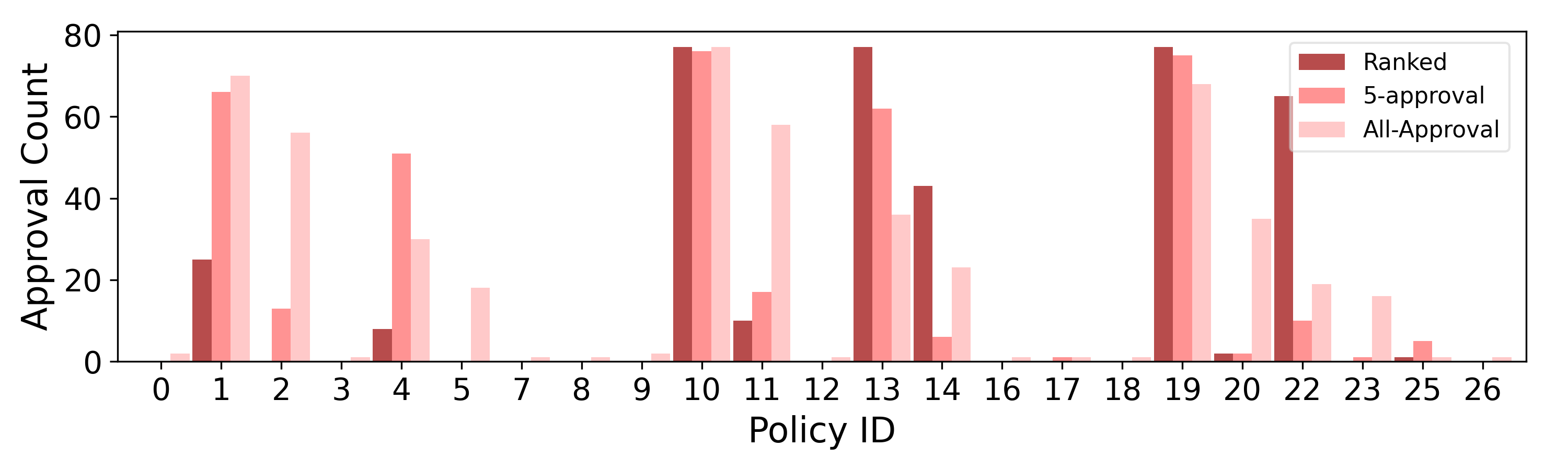}
    \caption{Voting counts of different voting types (single round).}
    \label{fig:vote_type}
\end{figure}

Figure~\ref{fig:vote_type} shows the distribution of votes from a single round obtained using each of the three voting methods. Overall, the results reveal a broadly consistent preference for a small set of top-ranked policies, particularly \textit{Policies~10, 13, and 19}. Using the instant-runoff procedure, \textit{Policy~10} emerges as the winner in the Ranked-Choice voting scheme. It also receives the highest number of votes in both the 5-approval and All-Approval settings, making it the winning policy across all three methods. While \textit{Policy~10} is neither the \textit{Utilitarian} nor the \textit{Egalitarian} optimum identified by the model, it lies close to the Pareto frontier (see Figure~\ref{fig:util_comparison_chicago}), suggesting that the simulated referendum is capable of selecting policies that align well with model-based recommendations.

The approval-based methods (5-Approval and All-Approval) produced more dispersed voting patterns, showing a greater tendency to favor policies with lower tax rates beyond the commonly agreed-upon options. The All-Approval method, in particular, admits several policies (e.g., \textit{Policies~5, 7, 8, 9, 16, 18, and 26}) that are not selected under either the Ranked-Choice or 5-Approval methods. Notably, these outlier policies are neither efficient nor equitable according to the model-based evaluation in Figure~\ref{fig:util_comparison_chicago}.

\begin{table}[h!]
\centering
\small
\caption{Winning policies, mean policy values and entropy (single round).}
\small
\begin{tabular}{lcccccccc}
\hline
 & \multicolumn{4}{c}{Winning Policy} & \multicolumn{3}{c}{Mean policy values} &  {Entropy} \\
 &ID&  $t$ & $r$ & $\tau$ & $t$ & $r$ & $\tau$  & $E$\\
\hline
Ranked-Choice & 10 & 1.000 & 0.750 & 0.500 & 0.833 & 0.917 & 0.750 & 2.739 \\
5-Approval &10  & 1.000 & 0.750 & 0.500 & 1.077 & 1.096 & 0.731 & 2.928  \\
All-Approval &10 &  1.000 & 0.750 & 0.500 & 0.978 & 1.185 & 0.587 & 3.565  \\
\hline
\end{tabular}
\label{tab:voting_results}
\end{table}

Table~\ref{tab:voting_results} summarizes the policy parameters of the winning policy and the mean policy values derived from each voting method. As noted earlier, \emph{Policy~10} emerges as the winner across all three methods. When comparing its parameter values with the corresponding mean policy values, we observe a clear pattern: while the average voter supports a comparable tax rate, they also express preference for significantly higher fare and driver fees than those set in the winning policy. This reflects a form of democratic compromise—although the public may lean toward user-based funding mechanisms in aggregate, the winning policy settles on more moderate fees to secure broader support.

\emph{Policy~19} consistently ranks as the top favored policy across all three voting methods. Notably, this policy coincides with the \textit{Utilitarian} policy identified by the model. In contrast, the \textit{Egalitarian} policy, \emph{Policy~20}, receives substantially fewer votes—failing to make the top five even under the All-Approval method. This result suggests that the aggregate preferences captured by GPT-4o agents in the Chicago simulation tend to align more closely with utilitarian values than with egalitarian ones.

Overall, we find strong consistency across the three voting mechanisms, with Ranked-Choice and 5-approval producing  similar outcomes. All-Approval voting, on the other hand, yields more scattered results and diverges notably from the other two. For this reason, we focus on the Ranked-Choice method in the remainder of our analysis.


\subsubsection{Impact of contextual information} \label{sec:res_models}

In this section, we run simulated referendums using three types of agents, each with different contextual information (see Section~\ref{sec:LLMExperiment}), namely,


\begin{enumerate}
    \item[1.] Chicago referendums carried out by \textbf{community-based agents}, who only use the knowledge acquired in pretraining, are labeled CHI-com. 
    \item[2.]  Chicago referendums carried out by \textbf{knowledge-augmented agents}, who are  prompted using localized data, are labeled CHI-know.
    \item[3.]   Chicago referendums carried out by \textbf{city-average agents}, who represent generic “average” residents of the city, are labeled CHI-avg.
\end{enumerate}
For each scenario, we run the simulation for 10 rounds, identify the winner for each round and analyze the distribution of votes across policies and the three policy levers. 

\begin{table}[ht!]
\centering
\small
\caption{Winning policy, mean entropy values and mean policy values across ten-round referendums. }
\small
\begin{tabular}{ll|l|c|c c c}
\hline
\textbf{Model} & \textbf{Scenario} & \textbf{Winner (counts)} & \textbf{Entropy $\bar{E}$} & \textbf{$\bar{t}_1$ ($\bar{e}_{t|1}$)} & \textbf{$\bar{r}_1$ ($\bar{e}_{r|1}$)} & \textbf{$\bar{\tau}_1$ ($\bar{e}_{\tau|1}$)} \\
\hline
    & CHI-com  & P10 (10)                         & 2.739 & 0.983 (0.21) & 0.782 (0.34) & 0.511 (0.15) \\
GPT-4o & CHI-know & P10 (8), P11 (2)              & 2.804 & 0.999 (0.02) & 0.772 (0.25) & 0.721 (0.98) \\
    & CHI-avg  & P10 (8), P11 (2)              & 2.611 & 1.000 (0.00) & 0.750 (0.00) & 0.600 (0.72) \\
\hline
\multirow{2}{*}{Claude-3.5} 
  & CHI-com  & P1 (6), P10 (4)             & 4.022 & 0.802 (1.29) & 0.951 (1.08) & 0.546 (1.52) \\
  & CHI-know & P1 (8), P4 (2)              & 3.902 & 0.782 (1.34) & 0.986 (1.04) & 0.579 (1.28) \\
\hline
GPT-4o & HOU-com  & P2 (10)                          & 3.583 & 0.635 (0.92) & 0.785 (0.29) & 0.895 (0.74) \\
\hline
\end{tabular}
\center
\footnotesize \textit{Note:} $E\in[0,4.75]$ measures the concentration of votes across all policies; $\bar{e}_{x|s}\in[0, 1.58]$ measures the concentration of the rank $s$ votes on the three policy levers ($x$) across the three levels.   The smaller the entropy value, the higher the concentration.

\label{tab:winning_policy_entropy}
\end{table}





As shown in Table~\ref{tab:winning_policy_entropy}, GPT-4o under the CHI-com setting exhibits highly stable preferences. Across 10 simulated referendum rounds, \emph{Policy~10} consistently emerges as the winning option. The distribution of votes across all policies shows only modest variation, reflected in a moderate entropy value of $E = 2.739$. The mean policy values indicate a clear and consistent preference for low fares, medium tax rates, and medium driver fees.

When agents are augmented with external contextual knowledge (CHI-know), the preference landscape shifts slightly. The winning set expands to include higher driver fee options—most notably \emph{Policy~11}, which wins in 20\% of the rounds. This change is also evident in the rank-1 mean policy values, where the average driver fee increases by 50\% compared to CHI-com. The entropy value rises slightly, suggesting a more dispersed distribution of preferences when additional knowledge is available.

In the CHI-avg setting, where a single agent represents the average Chicagoan, the set and frequency of winning policies mirror those under CHI-know. Does this imply that accounting for community diversity and context matter little? Not necessarily. The entropy value for CHI-avg is noticeably lower, indicating more concentrated preferences. Most strikingly, all rank-1 votes in this scenario support policies with a medium tax rate and the lowest fare—resulting in zero entropy for those policy levers.

In sum, while all three scenarios broadly agree on the most preferred policies, community-specific simulations introduce greater diversity in preference expression --- more so when additional contextual information is provided, while the averaged-agent setting yields more concentrated but potentially less nuanced outcomes.

\subsection{GPT-4o vs. Calude-3.5}
\label{sec:gpt_vs_claude}

To highlight the performance differences in voting behavior across different LLMs, we compare the winning policies and voting patterns generated by GPT-4o and Claude-3.5 in the CHI-com scenario. We then try to explain the observed differences by applying  sentiment analysis to the rationale texts from both the CHI-com and CHI-know scenarios.
\subsubsection{Winning policies and voting patterns}
Table~\ref{tab:winning_policy_entropy} shows that, for CHI-com, the winning policy set expands from \textit{Policy~10} under GPT-4o to include both \textit{Policy~10} and \textit{Policy~1} under Claude-3.5. As a result, the winning probability of \textit{Policy~10} falls from 100\% to just 40\% when switching to Claude-3.5. The key distinction between these two policies lies in the sales tax rate: \textit{Policy~10} preserves the status quo at 1\%, while \textit{Policy~1} lowers it to 0.5\%. For CHI-know, the winning policies diverge entirely: Claude-3.5 selects \textit{P1} and \textit{P4}, while GPT-4o favors \textit{P10} and \textit{P11}. Their mean policy values also drift much further apart, underscoring that the two LLMs respond to local contexts in markedly different ways.

Compared to GPT-4o, Claude-3.5 exhibits significantly more dispersed voting behavior, as reflected in a roughly 50\% increase in the overall entropy measure. The entropy values for individual policy levers are even higher—for instance, the entropy of rank-1 votes for the driver fee is 1.52, indicating a near-uniform distribution and aggregate indifference.

\begin{figure}[ht!]
    \centering
    \begin{subfigure}[b]{0.48\textwidth}
        \centering
        \includegraphics[width=\textwidth]{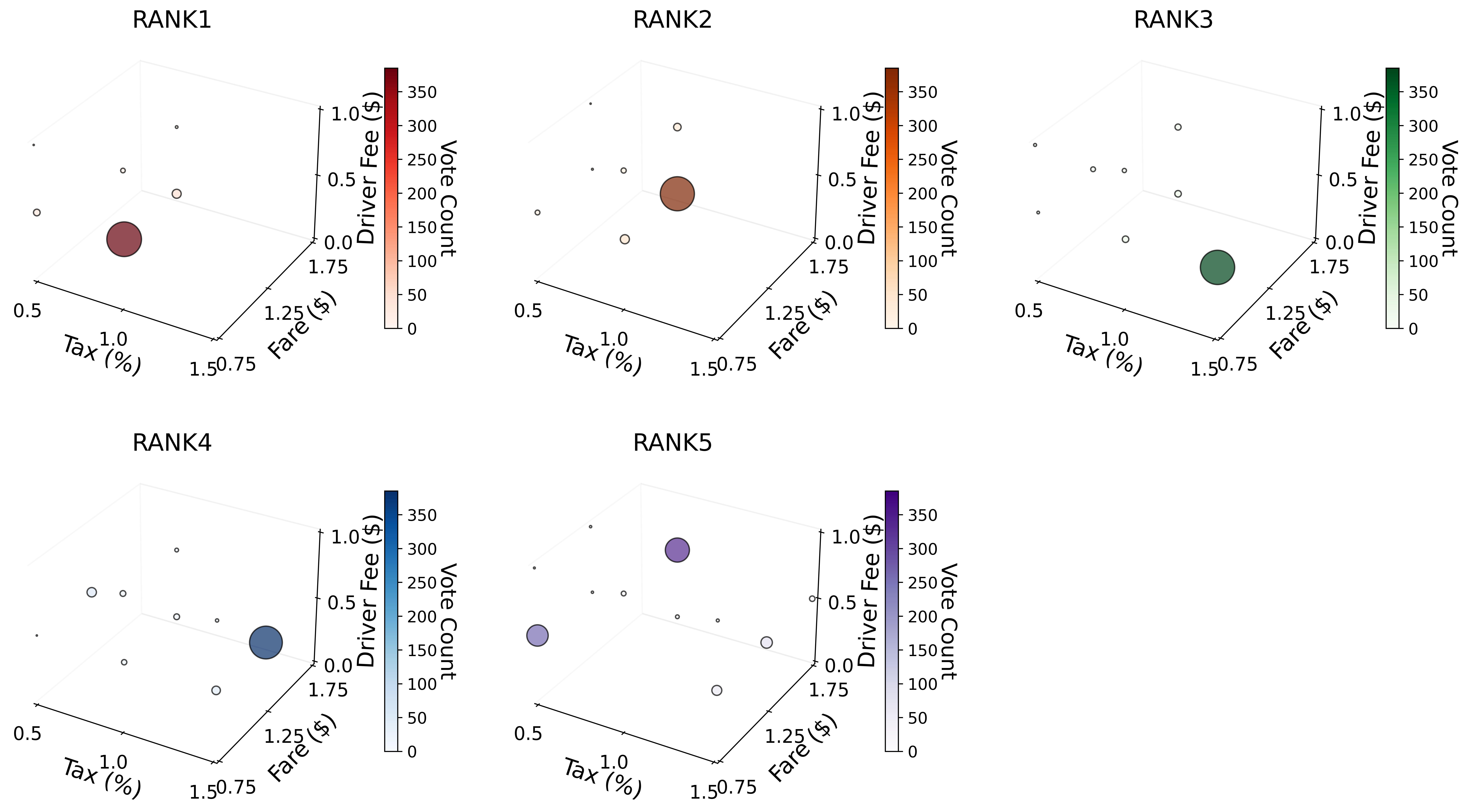}
        \caption{GPT-4o.}
        \label{fig:vote-gpt}
    \end{subfigure}
    \hfill
    \begin{subfigure}[b]{0.48\textwidth}
        \centering
        \includegraphics[width=\textwidth]{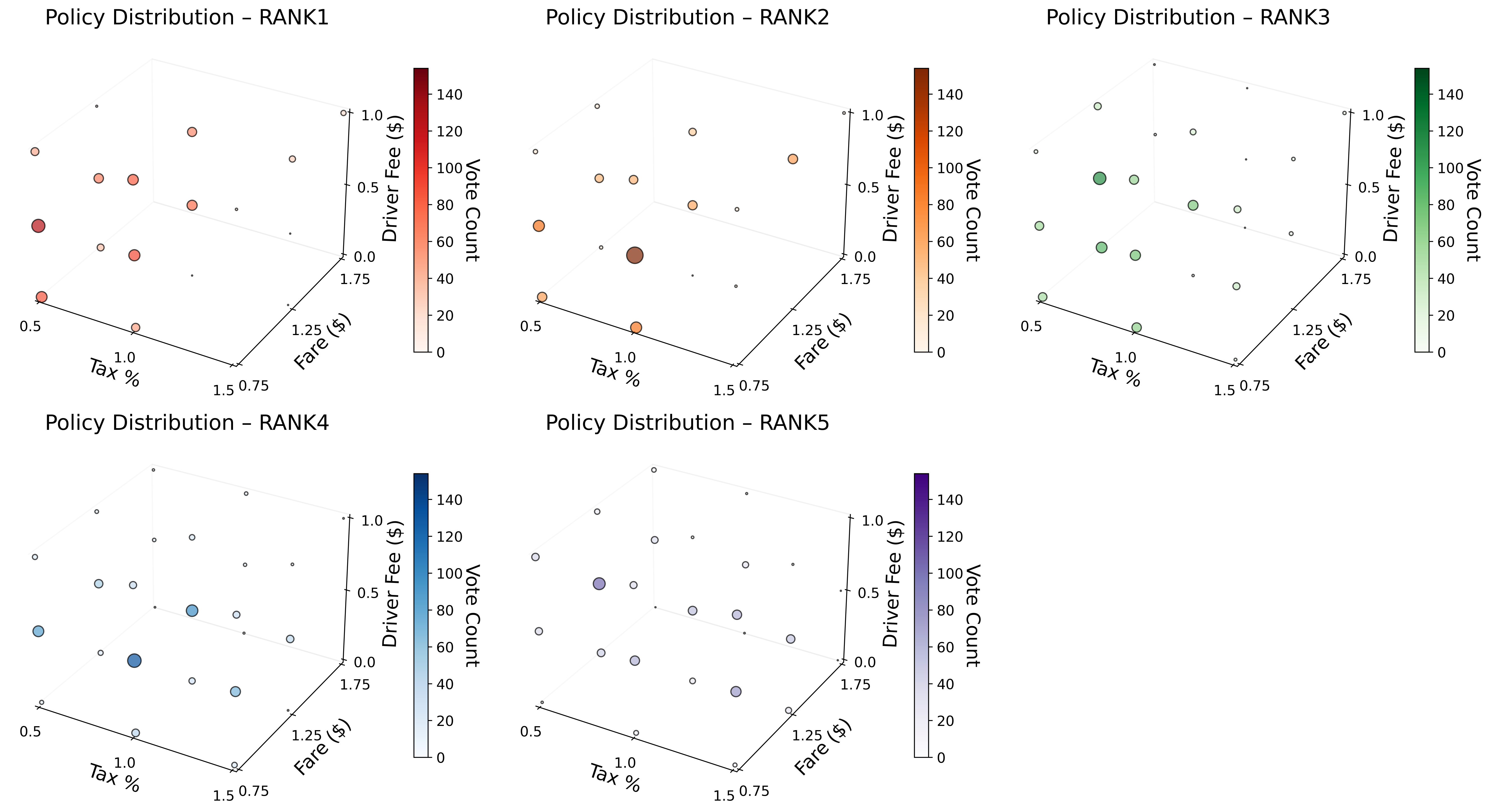}
        \caption{Claude-3.5.}
        \label{fig:vote-claude}
    \end{subfigure}

    \caption{Ranked-Choice voting outcomes generated by GPT-4o and Claude-3.5. Scenario CHI-com, ten rounds. 
    }
    \label{fig:vote-distribution-chi-com}    
\end{figure}

To further probe this divergence, Figure~\ref{fig:vote-distribution-chi-com} compares the full Ranked-Choice voting distributions generated in the referendums using GPT-4o and Claude-3.5. Each plot contains 27 lattice points representing the policy alternatives, with the size of each point proportional to the number of votes the policy received at a given rank across ten simulation rounds.
  
For CHI-com, the plots confirm that GPT-4o generates a much more concentrated voting distribution than Claude-3.5, as evidenced by the larger and more clearly defined clusters in Figure~\ref{fig:vote-gpt}. Despite the greater dispersion in Claude-3.5’s rankings, the average policy values associated with rank-1 votes remain similar across the two models, though Claude-3.5 shows a slight tilt toward policies with higher transit fares (see Table~\ref{tab:winning_policy_entropy}).

\begin{figure}[ht!]
    \centering
    \begin{subfigure}[b]{0.48\textwidth}
        \centering
        \includegraphics[width=\textwidth]{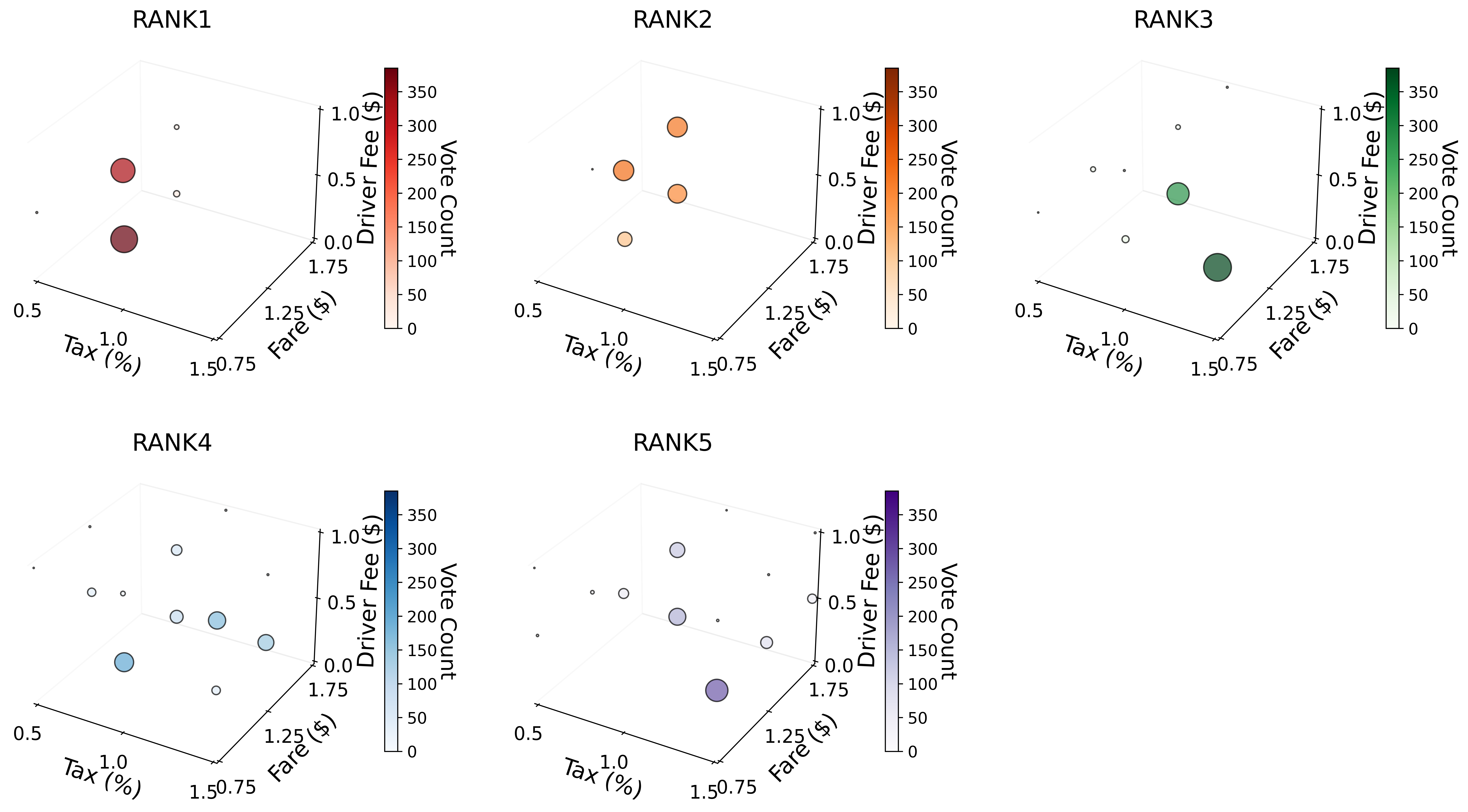}
        \caption{CHI-know GPT-4o}
        \label{fig:vote-gpt-chi-know}
    \end{subfigure}
    \hfill
    \begin{subfigure}[b]{0.48\textwidth}
        \centering
        \includegraphics[width=\textwidth]{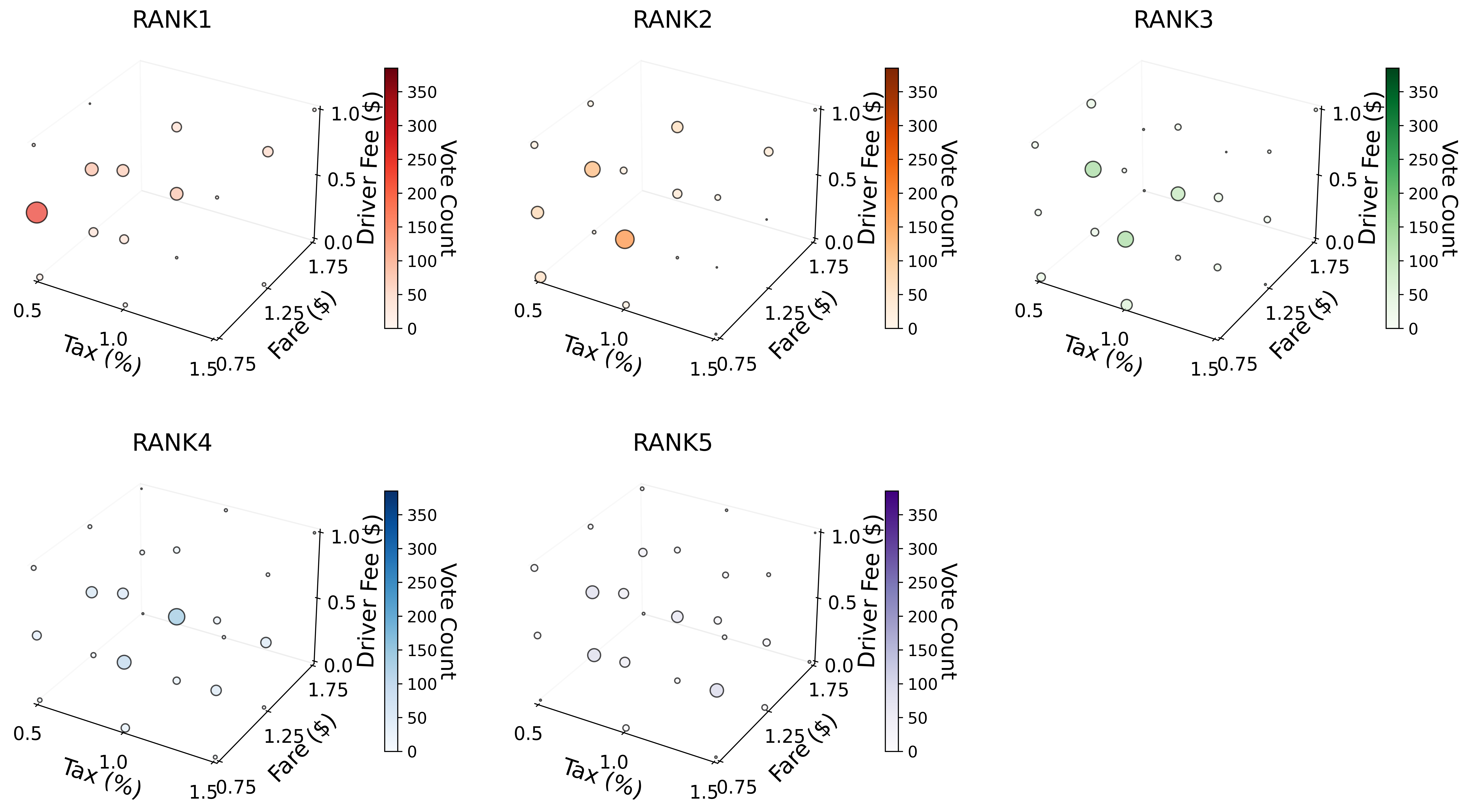}
        \caption{CHI-know Claude-3.5}
        \label{fig:vote-claude-chi-know}
    \end{subfigure}
    \caption{Ranked-Choice voting outcomes generated by GPT-4o and Claude-3.5. Scenario \textit{CHI-know}, ten rounds. 
    }

    \label{fig:vote-distribution-chi-know}
\end{figure}

As shown in Figure~\ref{fig:vote-distribution-chi-know}, providing additional contextual information disperses GPT-4o’s voting patterns, leading to consistently higher entropy values across the board (see Table~\ref{tab:winning_policy_entropy}). Interestingly, the effect is reversed for Claude-3.5. While less pronounced, the evidence—from the shift in winning policies (from a 4:6 split to a 2:8 split) and from the entropy values in Table~\ref{tab:winning_policy_entropy}—indicates that Claude-3.5’s voting patterns become more concentrated once contextual information is introduced.

\begin{figure}[ht!]
    \centering
    \begin{subfigure}[b]{0.48\linewidth}
        \centering
        \includegraphics[width=\linewidth]{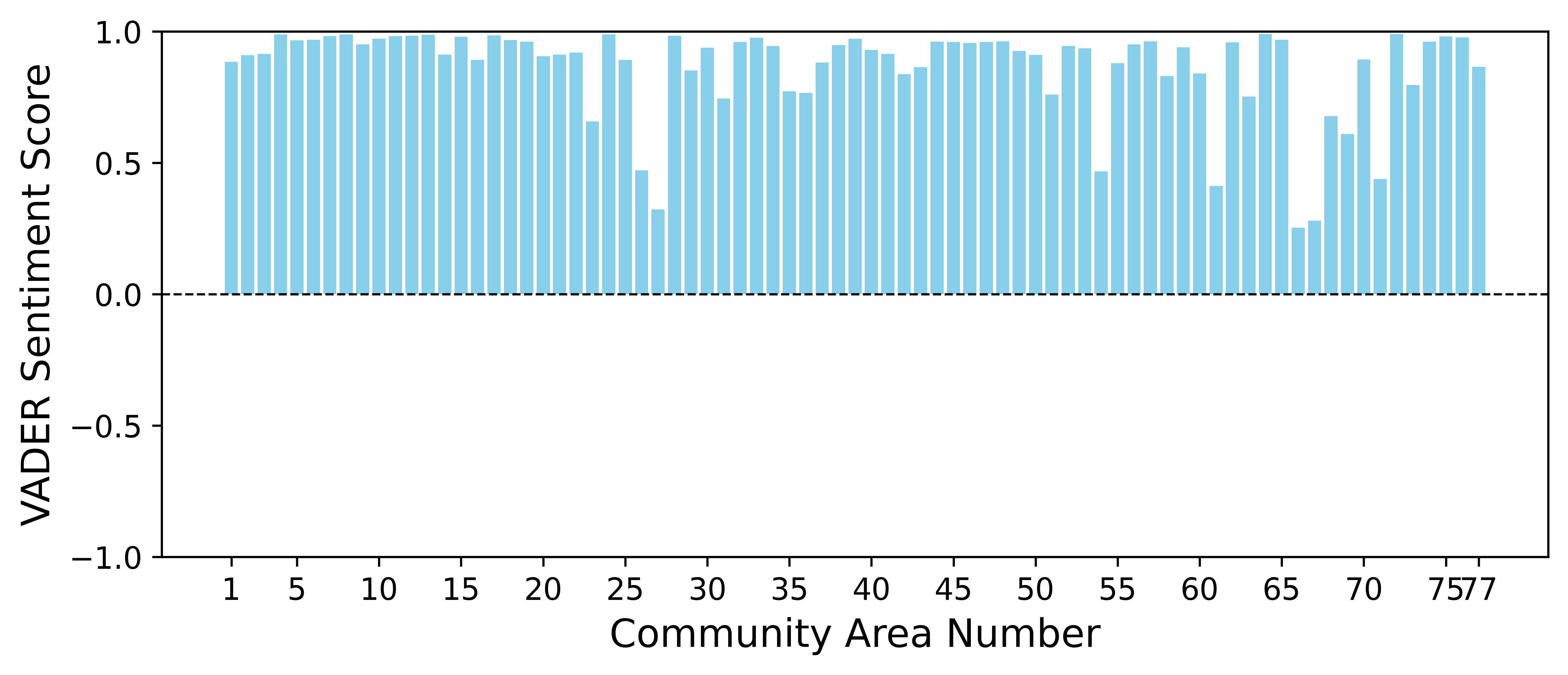}
        \caption{GPT-4o.}
        \label{fig:sentiment_chi_gpt}
    \end{subfigure}
    \hspace{0.02\linewidth} 
    \begin{subfigure}[b]{0.48\linewidth}
        \centering
        \includegraphics[width=\linewidth]{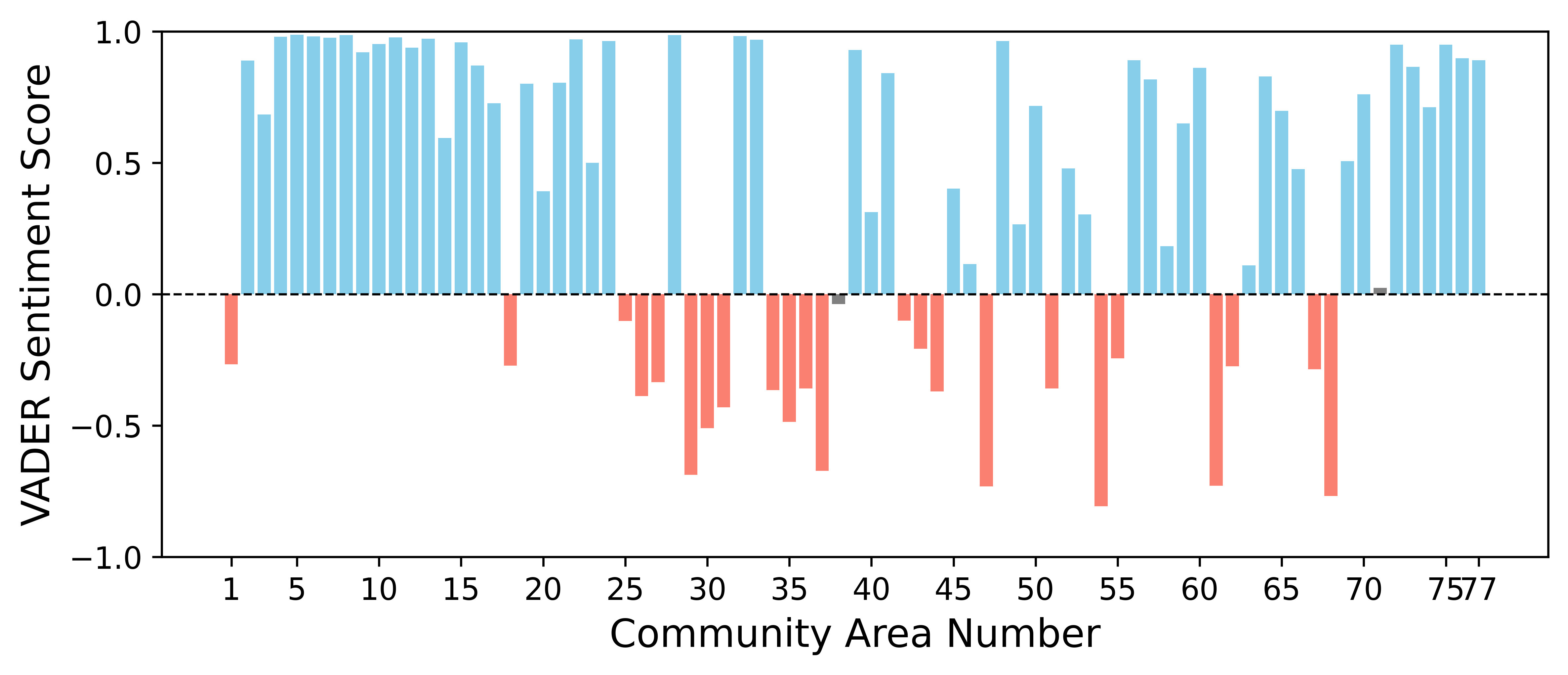}
        \caption{Claude-3.5.}
        \label{fig:sentiment_chi_claude}
    \end{subfigure}
    \caption{Sentiment scores of rationale texts for CHI-com. GPT-4o vs. Claude-3.5  across ten rounds. The scores range from -1 (perfectly negative tone) to 1 (perfectly positive tone). Each column corresponds to a distinct community area in Chicago, ordered by their official community area index used by the City of Chicago.}
    \label{fig:sentiment_chicago_both}
\end{figure}

\begin{figure}[h!]
    \centering
    \begin{subfigure}[b]{0.48\linewidth}
        \centering
        \includegraphics[width=\linewidth]{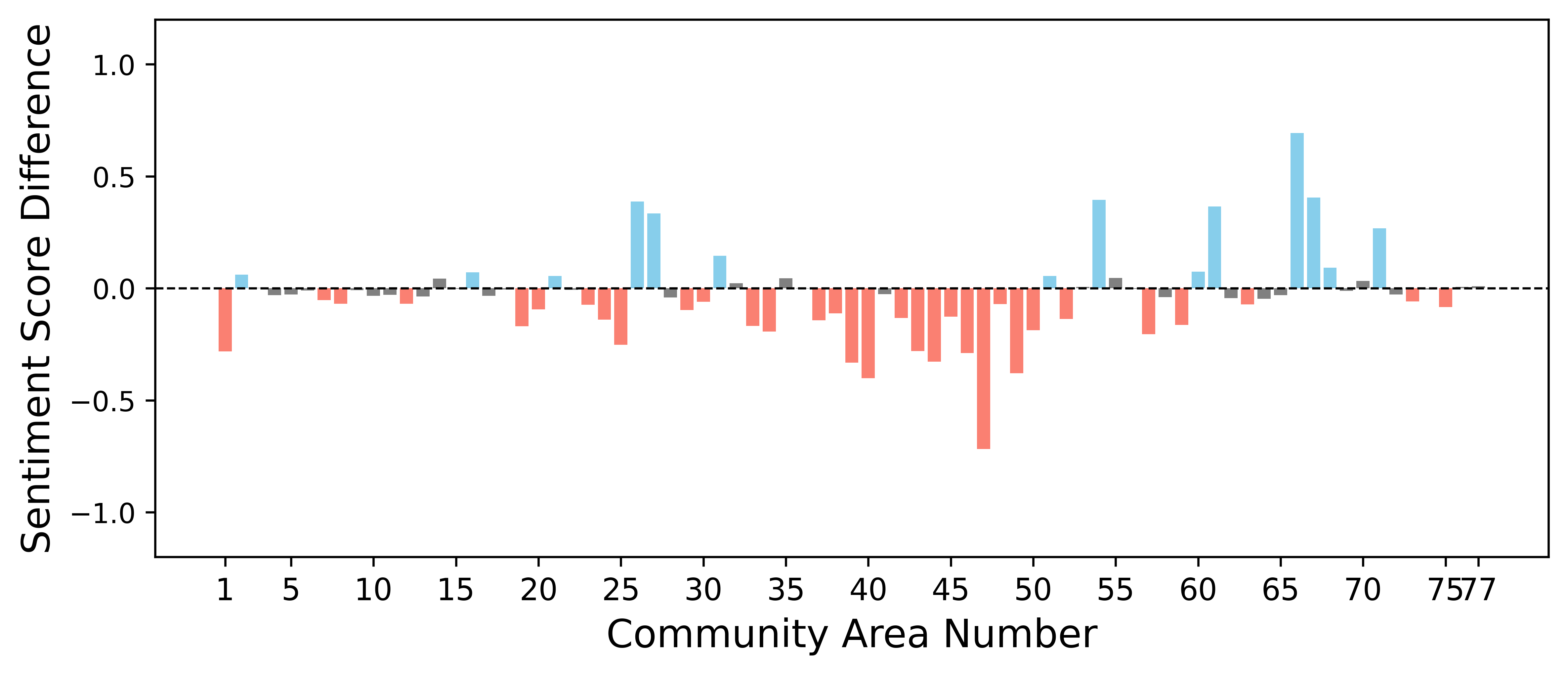}
        \caption{GPT-4o.}
        \label{fig:sentiment_chi_gpt_diff}
    \end{subfigure}
    \hspace{0.02\linewidth} 
    \begin{subfigure}[b]{0.48\linewidth}
        \centering
        \includegraphics[width=\linewidth]{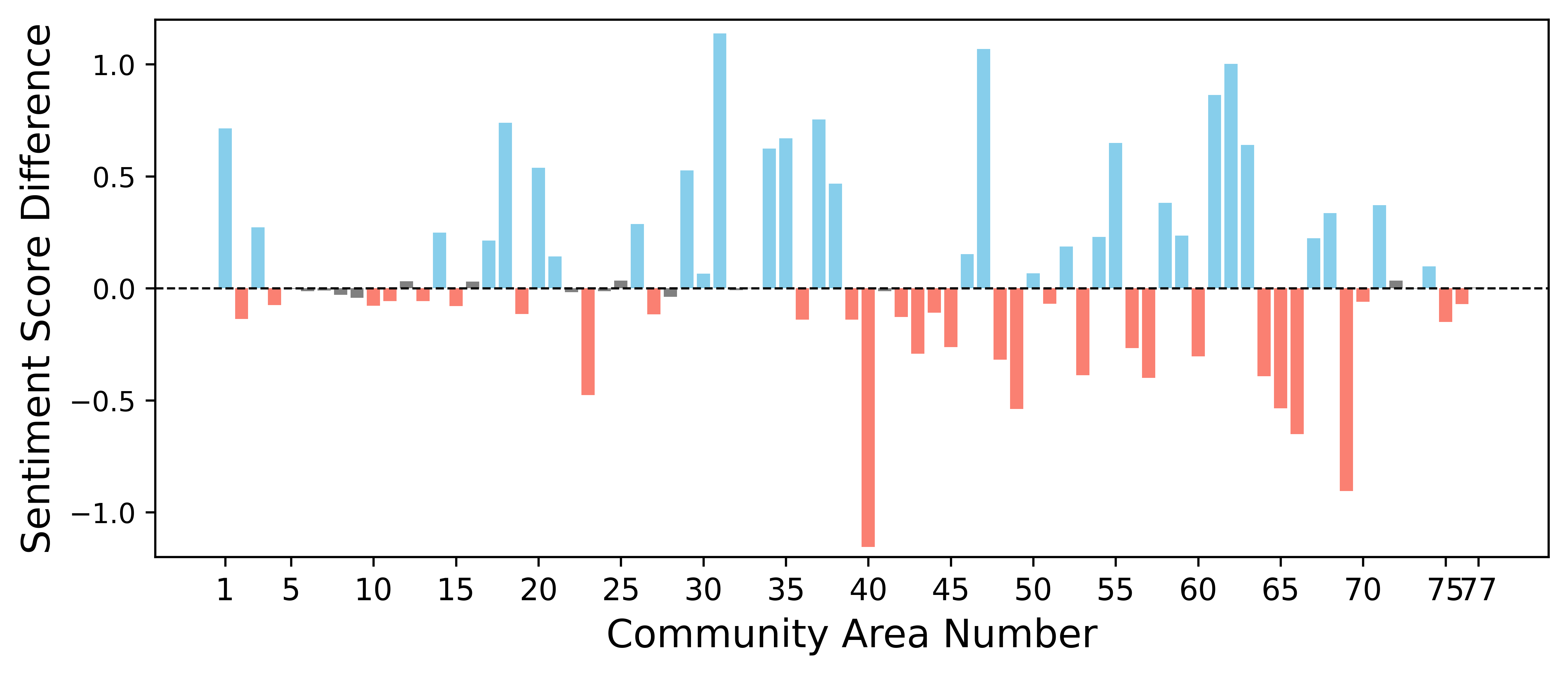}
        \caption{Claude-3.5.}
        \label{fig:sentiment_chi_claude_diff}
    \end{subfigure}
    \caption{Differences in the sentiment scores of rationale texts between CHI-know (with contextual information) and CHI-com (without). GPT-4o vs. Claude-3.5 agents across ten rounds.   Each column corresponds to a distinct community area in Chicago, ordered by their official community area index used by the City of Chicago.}
    \label{fig:sentiment_chicago_diff}
\end{figure}

\subsubsection{Sentiment analysis}
To better understand the divergent voting behaviors exhibited by the two LLMs—despite their broadly similar aggregate preferences—we turn to the sentiments expressed in the agents’ rationale texts. These rationales, generated as part of the structured response for each agent, provide insight into the tone and emotional framing of their decision-making. We process these texts using the VADER sentiment analyzer, which produces a compound score ranging from -1 (highly negative) to 1 (highly positive) for each community.

As shown in Figure~\ref{fig:sentiment_chicago_both}, GPT-4o agents exhibit consistently positive sentiment across all communities. In contrast, Claude-3.5 produces a much wider dispersion: while some agents are equally upbeat, a notable share—nearly one-third—express neutral or negative sentiment in their reasoning. This discrepancy is striking and may offer a partial explanation for the higher entropy and broader vote distribution observed with Claude-3.5.

In Figure~\ref{fig:sentiment_chicago_diff}, we compare the differences in sentiment scores between the CHI-know and CHI-com scenarios. The results highlight distinct ways in which the two models respond to contextual information. For GPT-4o, the information tends to moderate overly optimistic tones,  pushing sentiment scores downward. While corrections are frequent, the overall variation remains within a relatively narrow range. GPT-4o thus appears to internalize contextual information in a cautious manner, limiting drastic changes across community areas. In contrast, Claude-3.5 exhibits much stronger sensitivity to contextual inputs. The sentiment differences are both larger in magnitude (in some cases the swing exceeds 1.0) and more polarized, with sharp positive spikes in some community areas and pronounced negative shifts in others. 

While we cannot assert a direct causal link between sentiment and vote outcome, it is reasonable to hypothesize that sentiment affects how agents weigh trade-offs and interpret community needs. For example, a more pessimistic framing might lead agents to avoid policies with perceived downside risks, whereas a more optimistic tone could favor ambitious, redistributive strategies.

The more uniformly positive tone of GPT-4o raises further questions. Might this reflect a bias introduced during pretraining—perhaps an overexposure to institutional or promotional language that emphasizes uplift and resolution? Or could it indicate an underrepresentation of narratives from marginalized communities, which may affect how the model perceives hardship that some of the Chicago communities must have been experiencing? These questions, though beyond the scope of the present study, point to a broader concern: sentiment in LLM-generated reasoning may not merely reflect mood but encode deeper assumptions about the world. As such, it deserves closer scrutiny in future research on model alignment and fairness.

\subsection{Preference of LLM agents via regression}
\label{sec:res-preferences}
To delve deeper into the decision making rationale of LLM agents and unveil the impact of contextual information, we specify a set of regression models to reveal the relation between the agents' preferences for each of the three policy levers (tax, fare and driver fee) and their sociodemographic attributes. 


To quantify each agent's preference, we apply the Borda Count method, which assigns scores based on the rank order of each policy. This approach yields a numerical preference score for each policy in each community, averaged over 10 simulation rounds. Formally, let $\bar{s}_{ix}$ denote the average Borda score assigned according to the policy lever $x \in \mathcal{L}$ for community area $i$, i.e  
$$
\bar{s}_{ix} = \frac{1}{J}\sum_{j=1}^{J}\frac{\sum_{s=1}^{5}(6-s) q(x|k(i,j,s))}{\sum_{s=1}^{5}s}
$$
\noindent
where $j$ indexes the referendum round in the total set of $J$ rounds, and $s \in $$\{1,2,3,4,5\}$ denotes the rank position (with $s=1$ corresponding to the top rank). The term $\frac{6-s}{\sum_{s=1}^{5}s}$ specifies the normalized weight for the $s$-th rank, while $q(x|k(i,j,s))$ returns the level (i.e., $l, m$ or $h$) of policy lever $x$ in policy $k$ that is ranked by community $i$ at $s$ in round $j$. Note that three levels may be represented by different numerical values for different levers. 
Thus, $\bar{s}_{ix}$ is the average normalized Borda score assigned to policy lever $x$ by community $i$ across $J$ rounds under the ranked voting process.

In addition to standard sociodemographic attributes, we introduce interaction terms between each attribute and a binary treatment indicator \( D_i \), where \( D_i = 1 \) if the observation originates from the CHI-know condition (with factual context) and \( D_i = 0 \) if from the CHI-com condition (without additional facts). We specify an Ordinary Least Squares (OLS) regression model as follows:
$$
\bar{s}_{ix} = \beta_{x0} + \vbeta_{x}^\top \mathcal{X}_{i} + \gamma_{x 0} D_i+ \bm{\gamma}_x^\top \mathcal{X}_{i} D_{i} + \varepsilon_{i x},
$$
where \(\beta_{x 0}\) is the intersect associated with policy lever \(x\),  \( \mathcal{X}_{i} \) denotes the vector of sociodemographic covariates for agent \( i \) and \(\vbeta_x\) is the corresponding vector of coefficients. In addition, \(\gamma_{x0}\) is the coefficient for the treatment indicator \(D_i\), which captures the average effect (or fixed effect) of enriching LLM agents' knowledge with localized information.  The interaction coefficients vector \( \bm{\gamma}_x \) measures the marginal effects of providing factual information on the relation between given sociodemographic attributes and community preferences.  The sign of each element in \(\bm{\beta}_x\) indicates whether a higher covariate value increases or reduces the Borda score, whereas the sign of each element in \(\bm{\gamma}_x\) represents whether additional information further reinforces or weakens the relation between Borda score and sociodemographic covariates. 



We begin with a full set of available sociodemographic variables (e.g. income and race) and travel attributes (e.g., car ownership and access to transit). Using the voting results simulated by GPT-4o, we iteratively test the explanatory power of each variable. A  variable is retained if it
(i)  exhibits statistical significance according to standard t-tests and F-tests and
(ii) is not strongly correlated with one another. We also attempt to balance across different categories so that the selection includes at least one variable related to either race, car ownership, travel mode, or income.  As the final step, we compute the Variance Inflation Factors (VIFs) of the included variables to rule out the concern for multicollinearity.
Table~\ref{tab:summary_stats} reports descriptive statistics for the selected variables.  Since all variables are measured in percentages, they are all represented by a real number between 0 and 1 in regression to avoid the bias from scale differences.

\begin{table}[ht!]
\centering
\small
\caption{Summary statistics of sociodemographic and travel variables.}
\label{tab:summary_stats}
\begin{tabular}{l p{6.5cm} cccc}
\hline
\textbf{Variable} & \textbf{Description} & \textbf{Min} & \textbf{Max} & \textbf{Mean} & \textbf{Std} \\
\hline
non\_white    & Percentage of residents of non-White race & 16.00\% & 99.80\% & 72.77\% & 25.80\% \\
car\_no            & Percentage of households without cars  & 3.70\% & 58.00\% & 24.12\% & 12.92\% \\
tvl\_transit       & Percentage of commuters traveling by public transit & 4.20\% & 36.90\% & 19.50\% & 7.53\% \\
income\_less\_25k  & Percentage of households with annual income below \$25k & 6.10\% & 70.00\% & 23.19\% & 12.26\% \\
income\_150k\_plus & Percentage of households with annual income above \$150k & 0.60\% & 49.50\% & 17.21\% & 12.26\% \\
\hline
\end{tabular}
\end{table}





\begin{table}[ht!]
\centering
\small
\caption{Regression results for tax rate, transit fare, and driver fee under GPT-4o and Claude.}
\begin{tabular}{lll|ll|ll}
\hline
 & \multicolumn{2}{c|}{\textbf{Tax rate}} & \multicolumn{2}{c|}{\textbf{Transit fare}} & \multicolumn{2}{c}{\textbf{Driver fee}} \\
 & GPT-4o & Claude-3.5 & GPT-4o & Claude-3.5 & GPT-4o & Claude-3.5 \\
\hline
constant                 & 1.123$^{***}$ & 0.868$^{***}$ & 0.968$^{***}$ & 0.963$^{***}$ & 0.535$^{***}$ & 0.522$^{***}$ \\
non\_white           & 0.004    & -0.031   & 0.005    & 0.010    & -0.003   & -0.085$^{*}$  \\
car\_no                   & -0.009   & 0.073$^{**}$  & 0.008    & 0.018    & 0.024$^{**}$  & 0.032    \\
tvl\_transit              & 0.010$^{\dagger}$ & 0.025    & -0.002   & -0.059$^{***}$& -0.008   & 0.036    \\
income\_less\_25k         & -0.033$^{***}$& -0.056$^{*}$  & -0.013$^{\dagger}$  & 0.001    & -0.008   & 0.010    \\
income\_150k\_plus        & -0.013   & 0.076$^{***}$ & 0.021$^{**}$  & 0.157$^{***}$ & 0.024$^{**}$  & 0.111$^{*}$   \\
info\_dummy               & -0.025$^{***}$& -0.018   & -0.047$^{***}$& 0.037$^{**}$  & 0.150$^{***}$ & 0.024    \\
pct\_non\_white $\times$ info & 0.007    & 0.017    & 0.001    & -0.005   & 0.008    & 0.086$^{\dagger}$   \\
car\_no $\times$ info     & 0.000    & -0.017   & 0.003    & -0.048$^{\dagger}$  & 0.001    & 0.011    \\
tvl\_transit $\times$ info& -0.009   & 0.004    & -0.030$^{***}$& 0.026    & 0.025$^{**}$  & 0.032    \\
income\_less\_25k $\times$ info & 0.035$^{**}$  & 0.020    & -0.012   & -0.015   & 0.014    & -0.037   \\
income\_150k\_plus $\times$ info & -0.006   & 0.045    & 0.015    & -0.067$^{*}$  & 0.018    & 0.035    \\
\hline
R$^2$                     & 0.45     & 0.724    & 0.729    & 0.782    & 0.875    & 0.593    \\
Adj.\ R$^2$               & 0.407    & 0.702    & 0.708    & 0.765    & 0.865    & 0.561    \\
F-statistics              & 10.54 $^{***}$    & 33.81$^{***}$     & 34.79$^{***}$     & 46.18$^{***}$     & 89.99$^{***}$     & 18.79$^{***}$     \\
AIC                       & -582.6   & -276     & -608.3   & -332.4   & -596.2   & -131.6   \\
\hline
\end{tabular}
\vspace{0.3em}

\centering
\footnotesize \textit{Note.} Significance: $^{\dagger}p<0.1$, $^{*}p<0.05$, $^{**}p<0.01$, $^{***}p<0.001$.

\label{tab:regression_vote}
\end{table}

The final regression results  are reported in Table~\ref{tab:regression_vote}. We can see that 
the effect of the information dummy is pronounced in the GPT-4o models. Across all three policy domains, the variable is highly significant, indicating that the provision of additional information systematically shapes preferences and improves explanatory power. Notably, the negative coefficients for tax rate and transit fare suggest that, when contextual signals  are incorporated, GPT-4o registers stronger aversion to financial burdens imposed by higher taxes or fares. By contrast, the positive and relatively large coefficient for driver fee implies that information makes the model more inclined to endorse this revenue instrument, reflecting a redistribution logic that shifts costs toward car users rather than vulnerable transit riders.
Claude-3.5, in comparison, shows weaker and less consistent responses to the information variable, suggesting that it internalizes contextual cues less directly. With local context, Claude-3.5 agents become more open to raising transit fare, while indifferent to the other two instruments.

From Table~\ref{tab:regression_vote}, the regression outcomes show several consistent patterns across GPT-4o and Claude-3.5, with some notable differences in explanatory focus.

The regression result with GPT-4o agents indicates that low-income populations would favor a lower tax rate, and this effect is reinforced when local information is introduced. Claude-3.5 highlights both  low-income and high-income communities, and the households with no car ownership. The wealthier-community would favor a higher tax rate, while low-income communities oppose them, but not significantly affected by contextual detail.

In GPT-4o, low-income populations are linked to a preference for lower fares, while high-income populations are supporting higher fares. With local context,  the impact of transit-dependency on fare is significantly weakened, though transit-dependency itself (measured by transit share) is insignificant. Claude-3.5 similarly highlights income, with high-income communities favoring higher fares and transit share exerting a negative effect on fare.  Moreover, local context weakens the impact of both income and car-ownership effects.

For the driver fee, both GPT-4o and Claude-3.5 suggest that high-income communities are more likely to support higher fees on drivers. Under GPT-4o, communities with higher levels of carlessness tend to favor raising driver fees as a way to improve transportation conditions, and the contextual information strengthens the effect of transit-dependency, which however is not a significant factor. In contrast, Claude-3.5 indicates that minority communities are inclined to oppose higher driver fees, and this effect is no longer evident with contextual information.

Overall, the regression model explains more variance in tax rate and transit fare outcomes in the data generated by Claude-3.5, while showing stronger explanatory power for driver fee outcomes in GPT-4o. Substantively, GPT-4o agents focus narrowly on income and car ownership as key determinants, whereas Claude-3.5 agents draw on a broader set of sociodemographic factors, including car ownership, transit share, and minority status. Reassuringly, the signs of the significant correlations align with intuitive expectations: low-income communities tend to resist higher taxes and fares, while car-less or transit-reliant communities support higher fees on drivers. 

\subsection{Chicago vs Houston}
\label{sec:res_houston}

To assess how urban context influences the voting behavior of LLM agents, we replicate our simulation experiments in Houston, TX. Like Chicago, Houston is a large metropolitan area with a comparable population size, comprising 88 super neighborhoods. However, its transportation infrastructure, travel behavior, and broader policy environment differ markedly. Houston is more car-oriented, with lower transit usage and historically less investment in public transportation. These contrasts offer a useful testbed for evaluating the adaptability and generalization capacity of the LLM-based voting framework.

\begin{figure}[htbp]
    \centering
    \begin{subfigure}[b]{0.32\textwidth}
        \centering
        \includegraphics[width=\textwidth]{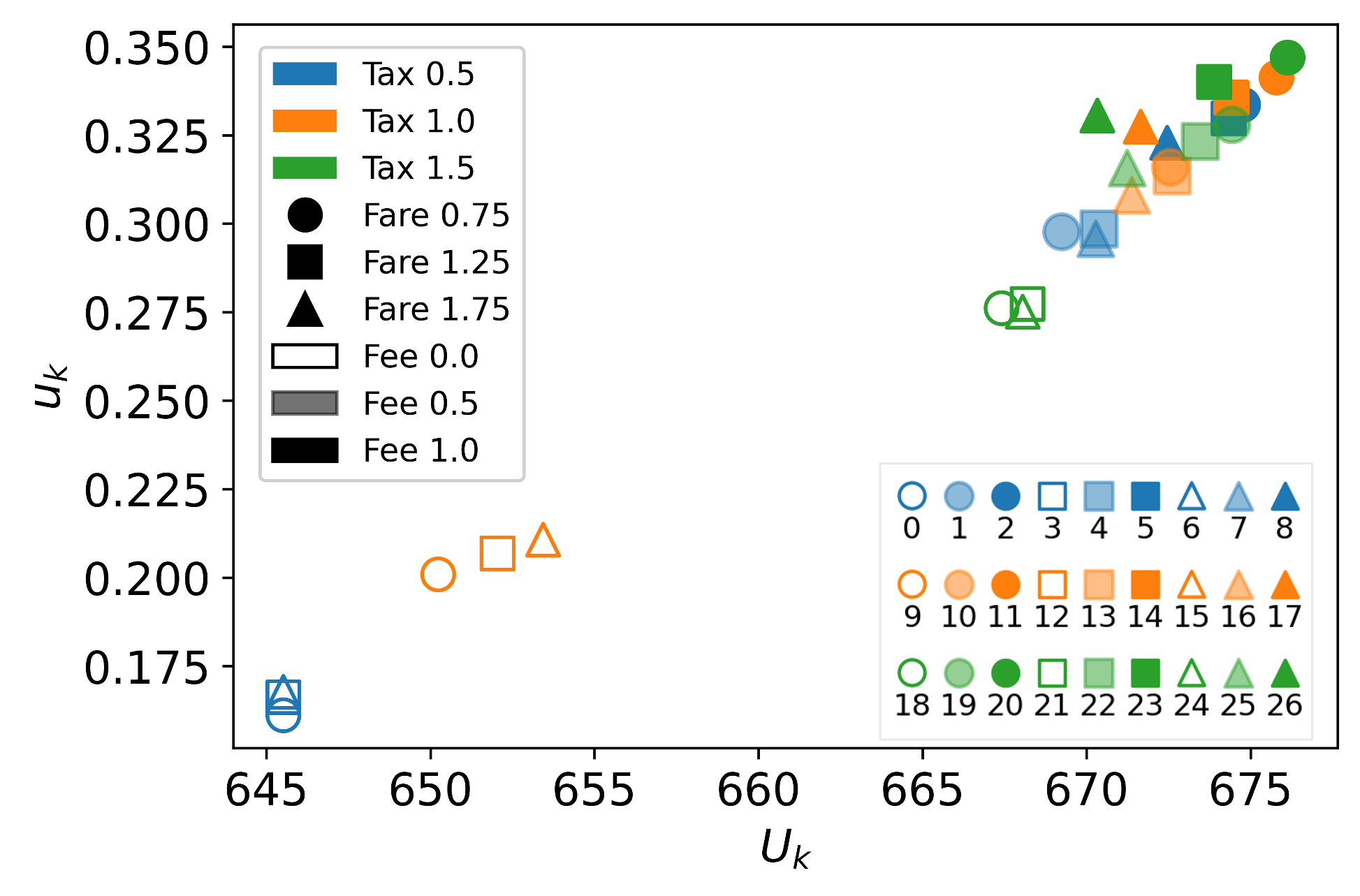}
        \label{fig:util-houston}
        \vspace{-10mm}
    \end{subfigure}
    \hfill
    \begin{subfigure}[b]{0.32\textwidth}
        \centering
        \includegraphics[width=\textwidth]{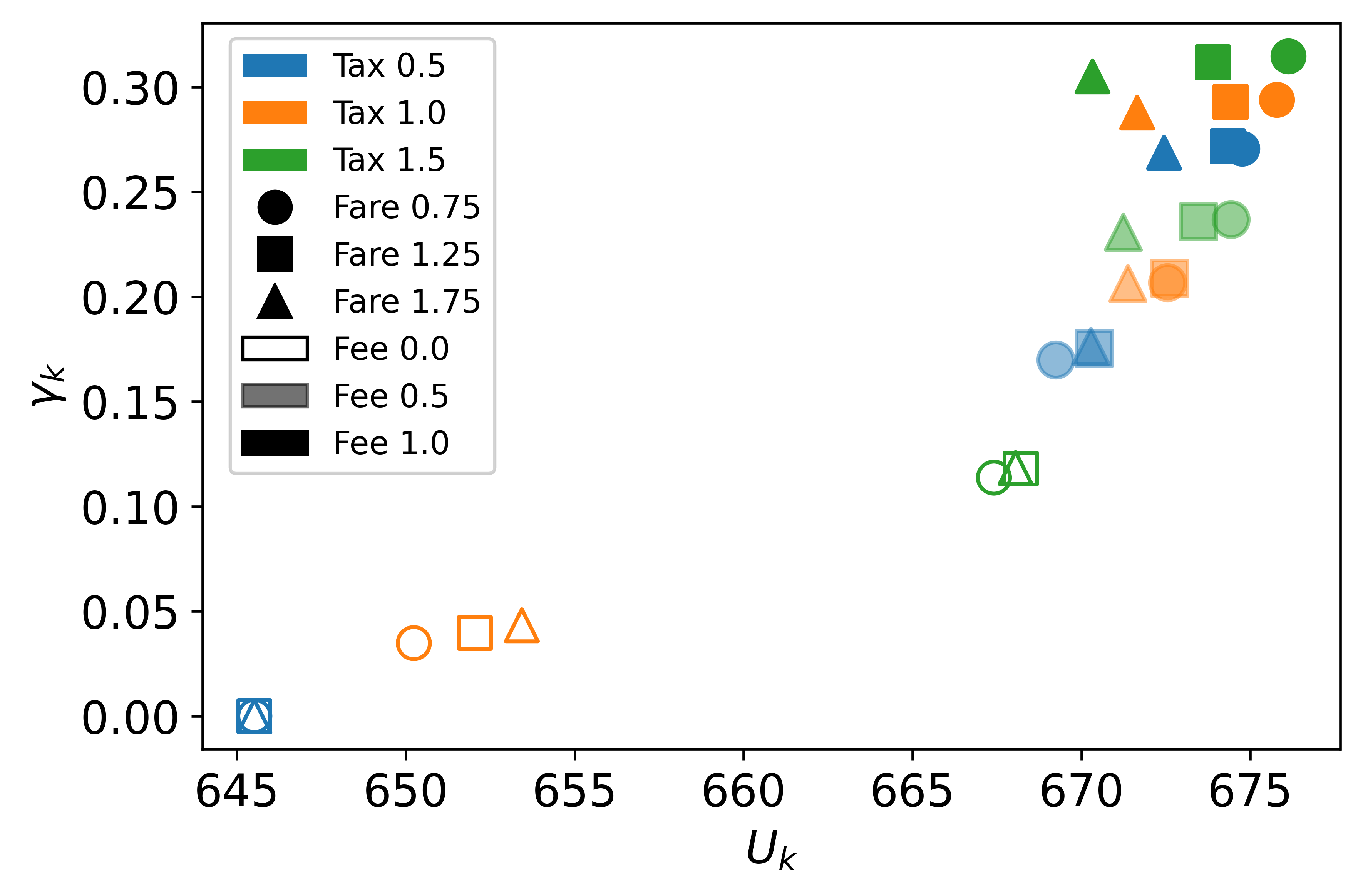}
        \label{fig:transit-hou}
        \vspace{-10mm}
    \end{subfigure}
    \hfill
    \begin{subfigure}[b]{0.32\textwidth}
        \centering
        \includegraphics[width=\textwidth]{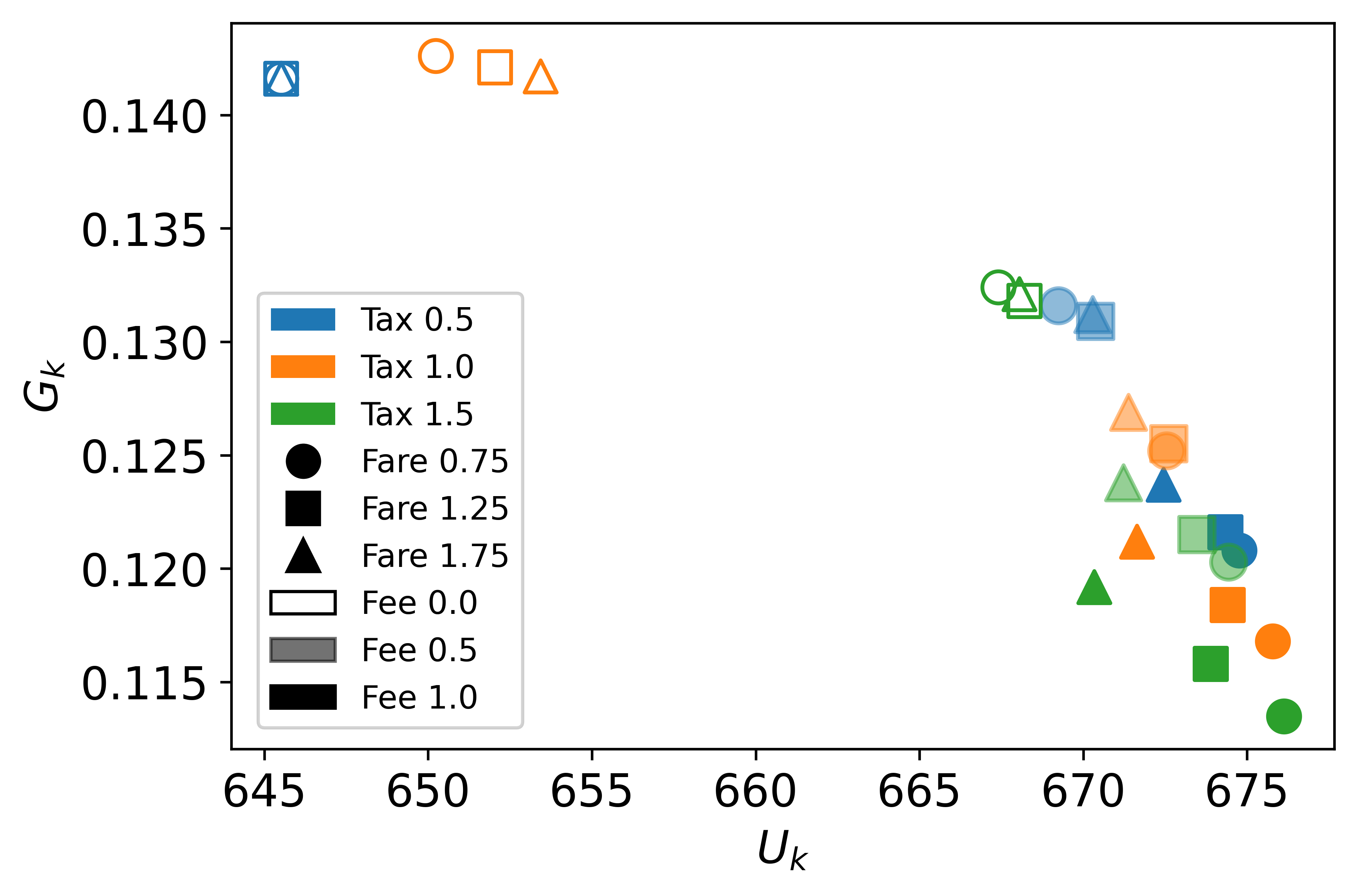}
        \label{fig:gini-hou}
        \vspace{-10mm}
    \end{subfigure}
    \caption{Comparison of $u_k$, $\gamma_k$, and $G_k$ against $U_k$ for different policy configurations in Houston.}
    \label{fig:houston_metrics}
\end{figure}

Figure~\ref{fig:houston_metrics} presents the model-based evaluation of the 27 policy proposals for Houston. Notably, \textit{Policy 20} emerges as the clear Pareto-optimal choice—it lies at the frontier in all three metrics considered: total utility ($U_k$), minimum utility ($u_k$), and the Gini index ($G_k$).  Compared to the common winner in Chicago (\textit{Policy 10}), \textit{Policy 20} maintains the same transit fare but sets both the sales tax and driver fee at their highest levels (1.5\% and \$1, respectively).

This result is consistent with the model’s logic. Given Houston’s low transit share and poor service levels, there is significant potential for improvement. Raising additional revenue through broader taxation and congestion-based fees—while keeping fares affordable—is  the most effective way to boost transit access and system efficiency. 

The referendums simulated by GPT-4o in Houston consistently select \textit{Policy 2} as the winning policy across all ten rounds. Compared to \textit{Policy 20}—the model’s optimal choice—the only point of disagreement is the sales tax rate. While the model advocates raising the tax to its maximum level (1.5\%), the GPT-4o agents favor keeping it at the minimum (0.5\%).

A closer comparison between \textit{Policy 2} (the voter-endorsed choice in Houston) and \textit{Policy 10} (the winning policy in Chicago) suggests that GPT-4o agents in Houston are less favorable toward taxation and more receptive to driver fees than their counterparts in Chicago. Moreover, the entropy values reported in Table~\ref{tab:winning_policy_entropy}, which show that the voting distribution in Houston exhibits significantly higher entropy, indicate greater dispersion and diversity of preferences among agents.

While these findings remain preliminary, they offer intriguing clues about the LLM’s capacity to reflect contextual variation. The divergence in voting patterns may signal an implicit sensitivity to the political and cultural ethos of different urban environments—an area that warrants further investigation.


\section{Conclusion}
Traditional travel forecasting and decision models often recommend solutions that appear sound in theory but falter under public scrutiny. Existing methods of gathering community input—surveys, hearings, and focus groups—are slow, costly, and disconnected from the optimization process. The result is a persistent misalignment between policy recommendations and what communities are willing to support. This study explores whether LLMs, operating as autonomous agents in simulated citywide referendums, can provide timely insights into public support and policy preferences. By embedding agents within a realistic decision framework and prompting them with localized context and performance metrics, we observed their collective voting patterns and inferred their underlying priorities. Our simulation experiments examined multiple voting methods, LLM types, and urban contexts. The key findings are summarized as follows:
\begin{itemize}
\item[1.] \textit{Alignment and divergence with model-based benchmarks}: By and large, LLM agents selected policies that, while differing in detail, reflected broadly similar priorities to the model-based Pareto-optimal solutions. Notably, LLMs consistently showed a stronger preference for lower tax rates than those prescribed by the model.
\item[2.] \textit{Voting method robustness}: Across Ranked-Choice, 5-Approval, and All-Approval schemes, the top-ranked policies remained largely stable, although approval-based methods introduced more variability at the margins. This indicates that LLM referenda outcomes were not highly sensitive to the voting mechanism.
\item[3.] \textit{Impact of model choice}: GPT-4o generated more consistent and concentrated voting patterns, whereas Claude-3.5 produced more dispersed responses. Despite these behavioral differences, both models converged on broadly similar average policy preferences.
\item[4.] \textit{Sentiment as an explanatory lens}: Sentiment analysis of agent rationales showed that GPT-4o agents expressed uniformly positive sentiments, while Claude-3.5 agents varied more in tone across contexts. These affective differences may contribute to the decisiveness or dispersion of voting outcomes, though further research is needed to clarify causal mechanisms.
\item[5.] \textit{Determinants of voting behavior}: Regression analysis shows that  GPT-4o agents rely more narrowly on income and car ownership as key determinants, whereas Claude-3.5 agents incorporate a broader range of sociodemographic factors, including car ownership, transit share, and minority status. Importantly, the signs of significant correlations align with intuitive expectations: low-income communities tend to resist higher taxes and fares, while car-less or transit-reliant communities support higher fees on drivers.  
\item[6.] \textit{Context sensitivity and generalization}: When applied to Houston—a car-centric city with a distinct transit landscape and sociopolitical context—GPT-4o agents favored a different policy (\textit{Policy 2}) than in Chicago (\textit{Policy 10}). These preferences reflected lower tolerance for taxation and higher acceptance of driver fees, along with greater entropy, suggesting an implicit sensitivity to contextual and cultural variation across urban settings.
\end{itemize}

The observed variation in LLM behavior across urban contexts and model types—especially in voting patterns and sentiment expression—raises several questions that warrant closer examination in future research.

First, the divergence in sentiment profiles between GPT-4o and Claude-3.5 highlights not just stylistic differences, but model-level attitudes that are embedded during pretraining. GPT-4o’s tendency toward uniformly optimistic or compliant responses, for instance, is consistent with known alignment choices made to produce more agreeable user-facing outputs. Such tendencies go beyond task prompts and reflect structural biases in how models are tuned for interaction. Addressing these issues will likely require interventions at the model design and alignment stage, rather than adjustments at the user level.

Second, while this study treated LLM outputs as approximations of human preferences under structured guidance, we did not examine how prompt engineering, framing effects, or representational biases might influence simulation outcomes. Future work could vary prompt structures and persona designs to test sensitivity to framing effects, detect anchoring biases, and evaluate whether certain communities or policy perspectives are systematically amplified or marginalized.

Third, our results indicate that LLMs may encode broad contextual awareness—for example, capturing political or cultural differences between cities—but the degree to which this awareness aligns with empirical reality remains uncertain. Future research could connect LLM outputs with survey or ethnographic data to assess fidelity to actual community preferences. Similarly, sentiment measures could be validated against local indicators such as social stress, economic hardship, or public satisfaction with transit services.

At their core, LLMs are not only decision aids or behavioral simulators but also repositories of embedded social priors. As these models increasingly participate in policy deliberation, understanding—and, where necessary, correcting—those priors will be critical to safeguarding democratic legitimacy, procedural fairness, and value alignment in AI-assisted public planning.

\newpage
\bibliography{library}

\newpage
\appendix

\section{Available policy package sets}

\begin{table}[htbp]
\centering
\caption{Policy and their model-based performance metrics for Chicago.}
\begin{adjustbox}{max width=\textwidth}
\begin{tabular}{cccccccccccc}
\toprule
ID & tax rate & transit fare & driver fee & drive\_time & bus\_time & drive\_cost & bus\_cost & Transit\% & Utotal & Umin & Gini \\
\midrule
0  & 0.5 & 0.75 & 0   & 25.5  & 68.72 & 5.43 & 0.75 & 12.51 & 865.0972 & 0.2093 & 0.1544 \\
1  & 0.5 & 0.75 & 0.5 & 19.11 & 57.43 & 5.93 & 0.75 & 39.28 & 977.3733 & 0.3212 & 0.1086 \\
2  & 0.5 & 0.75 & 1   & 18.02 & 54.64 & 6.43 & 0.75 & 50.50 & 973.0646 & 0.3386 & 0.0957 \\
3  & 0.5 & 1.25 & 0   & 23.16 & 64.12 & 5.43 & 1.25 & 20.98 & 911.2723 & 0.2552 & 0.1365 \\
4  & 0.5 & 1.25 & 0.5 & 19.01 & 54.51 & 5.93 & 1.25 & 40.49 & 978.7941 & 0.3153 & 0.1096 \\
5  & 0.5 & 1.25 & 1   & 17.97 & 51.99 & 6.43 & 1.25 & 50.98 & 971.9173 & 0.3287 & 0.0991 \\
6  & 0.5 & 1.75 & 0   & 22.40 & 60.33 & 5.43 & 1.75 & 23.71 & 926.1360 & 0.2617 & 0.1326 \\
7  & 0.5 & 1.75 & 0.5 & 19.06 & 52.48 & 5.93 & 1.75 & 39.89 & 971.5802 & 0.3019 & 0.1149 \\
8  & 0.5 & 1.75 & 1   & 18.08 & 50.56 & 6.43 & 1.75 & 50.00 & 973.1771 & 0.3111 & 0.1071 \\
9  & 1.0 & 0.75 & 0   & 21.72 & 62.79 & 5.43 & 0.75 & 26.14 & 934.5490 & 0.2827 & 0.1257 \\
10 & 1.0 & 0.75 & 0.5 & 18.80 & 55.79 & 5.93 & 0.75 & 42.79 & 982.4324 & 0.3287 & 0.1027 \\
11 & 1.0 & 0.75 & 1   & 17.80 & 53.32 & 6.43 & 0.75 & 52.44 & 978.2246 & 0.3425 & 0.0917 \\
\textbf{12} & \textbf{1.0} & \textbf{1.25} & \textbf{0}   & \textbf{21.00} &\textbf{ 58.80} & \textbf{5.43} & \textbf{1.25} & \textbf{28.72} & \textbf{950.1400} & \textbf{0.2884} & \textbf{0.1219} \\
13 & 1.0 & 1.25 & 0.5 & 18.76 & 53.21 & 5.93 & 1.25 & 43.31 & 980.2506 & 0.3196 & 0.1056 \\
14 & 1.0 & 1.25 & 1   & 17.79 & 51.00 & 6.43 & 1.25 & 52.56 & 973.3634 & 0.3304 & 0.0964 \\
15 & 1.0 & 1.75 & 0   & 20.98 & 56.44 & 5.43 & 1.75 & 28.80 & 948.0086 & 0.2799 & 0.1244 \\
16 & 1.0 & 1.75 & 0.5 & 18.86 & 51.47 & 5.93 & 1.75 & 42.21 & 970.1432 & 0.3039 & 0.1123 \\
17 & 1.0 & 1.75 & 1   & 18.08 & 51.07 & 6.43 & 1.75 & 50.00 & 961.5373 & 0.3061 & 0.1089 \\
18 & 1.5 & 0.75 & 0   & 20.07 & 58.51 & 5.43 & 0.75 & 32.02 & 965.1986 & 0.3074 & 0.1142 \\
19 & 1.5 & 0.75 & 0.5 & 18.52 & 54.37 & 5.93 & 0.75 & 45.82 & 985.7996 & 0.3340 & 0.0981 \\
20 & 1.5 & 0.75 & 1   & 17.58 & 52.16 & 6.43 & 0.75 & 54.21 & 981.6979 & 0.3454 & 0.0883 \\
21 & 1.5 & 1.25 & 0   & 19.81 & 55.50 & 5.43 & 1.25 & 32.94 & 969.0859 & 0.3030 & 0.1150 \\
22 & 1.5 & 1.25 & 0.5 & 18.52 & 52.10 & 5.93 & 1.25 & 45.79 & 980.4515 & 0.3224 & 0.1025 \\
23 & 1.5 & 1.25 & 1   & 17.61 & 50.14 & 6.43 & 1.25 & 54.02 & 973.6848 & 0.3315 & 0.0942 \\
24 & 1.5 & 1.75 & 0   & 20.07 & 53.91 & 5.43 & 1.75 & 32.03 & 959.1524 & 0.2888 & 0.1201 \\
25 & 1.5 & 1.75 & 0.5 & 18.66 & 50.61 & 5.93 & 1.75 & 44.28 & 967.8360 & 0.3050 & 0.1102 \\
26 & 1.5 & 1.75 & 1   & 17.74 & 48.82 & 6.43 & 1.75 & 52.92 & 957.9162 & 0.3126 & 0.1029 \\
\bottomrule
\end{tabular}
\end{adjustbox}
\label{tab:policy_table}
\end{table}

\begin{table}[ht!]
\centering
\caption{Policy and their model-based performance metrics for Houston.}
\begin{adjustbox}{width=\textwidth}
\begin{tabular}{cccccccccccc}
\toprule
ID & tax rate & transit fare & driver fee & drive\_time & transit time & drive cost & transit cost & Transit \% & Utotal & Umin & Gini \\
\midrule
0  & 0.5 & 0.75 & 0   & 25.98 & 61.55 & 7.19 & 0.75 & 0.00   & 645.5176 & 0.1611 & 0.1416 \\
1  & 0.5 & 0.75 & 0.5 & 23.47 & 58.58 & 7.69 & 0.75 & 16.99  & 669.2392 & 0.2977 & 0.1316 \\
2  & 0.5 & 0.75 & 1   & 22.62 & 56.06 & 8.19 & 0.75 & 27.06  & 674.7613 & 0.3336 & 0.1208 \\
3  & 0.5 & 1.25 & 0   & 25.98 & 61.25 & 7.19 & 1.25 & 0.00   & 645.5176 & 0.1662 & 0.1416 \\
4  & 0.5 & 1.25 & 0.5 & 23.42 & 57.54 & 7.69 & 1.25 & 17.53  & 670.3823 & 0.2985 & 0.1309 \\
5  & 0.5 & 1.25 & 1   & 22.61 & 55.05 & 8.19 & 1.25 & 27.17  & 674.3346 & 0.3297 & 0.1216 \\
6  & 0.5 & 1.75 & 0   & 25.98 & 61.01 & 7.19 & 1.75 & 0.00   & 645.5176 & 0.1677 & 0.1416 \\
7  & 0.5 & 1.75 & 0.5 & 23.41 & 56.62 & 7.69 & 1.75 & 17.65  & 670.2816 & 0.2958 & 0.1312 \\
8  & 0.5 & 1.75 & 1   & 22.64 & 54.23 & 8.19 & 1.75 & 26.88  & 672.4483 & 0.3229 & 0.1237 \\
9  & 1.0 & 0.75 & 0   & 25.12 & 61.23 & 7.19 & 0.75 & 3.48   & 650.2375 & 0.2009 & 0.1426 \\
10 & 1.0 & 0.75 & 0.5 & 23.17 & 57.20 & 7.69 & 0.75 & 20.67  & 672.5462 & 0.3159 & 0.1252 \\
11 & 1.0 & 0.75 & 1   & 22.40 & 55.07 & 8.19 & 0.75 & 29.39  & 675.7829 & 0.3413 & 0.1168 \\
\textbf{12} & \textbf{1.0} &\textbf{ 1.25 }& \textbf{0}   & \textbf{25.00} & \textbf{60.77} & \textbf{7.19 }& \textbf{1.25} & \textbf{3.96}   & \textbf{652.0485} & \textbf{0.2068 }& \textbf{0.1421} \\
13 & 1.0 & 1.25 & 0.5 & 23.16 & 56.22 & 7.69 & 1.25 & 20.87  & 672.6134 & 0.3135 & 0.1255 \\
14 & 1.0 & 1.25 & 1   & 22.41 & 54.17 & 8.19 & 1.25 & 29.27  & 674.4099 & 0.3356 & 0.1184 \\
15 & 1.0 & 1.75 & 0   & 24.91 & 60.25 & 7.19 & 1.75 & 4.33   & 653.4359 & 0.2107 & 0.1417 \\
16 & 1.0 & 1.75 & 0.5 & 23.18 & 55.39 & 7.69 & 1.75 & 20.64  & 671.3721 & 0.3080 & 0.1269 \\
17 & 1.0 & 1.75 & 1   & 22.46 & 53.45 & 8.19 & 1.75 & 28.78  & 671.6410 & 0.3274 & 0.1212 \\
18 & 1.5 & 0.75 & 0   & 23.88 & 58.93 & 7.19 & 0.75 & 11.37  & 667.4096 & 0.2761 & 0.1324 \\
19 & 1.5 & 0.75 & 0.5 & 22.92 & 56.03 & 7.69 & 0.75 & 23.68  & 674.4277 & 0.3280 & 0.1203 \\
20 & 1.5 & 0.75 & 1   & 22.19 & 54.20 & 8.19 & 0.75 & 31.47  & 676.1273 & 0.3470 & 0.1135 \\
21 & 1.5 & 1.25 & 0   & 23.85 & 58.10 & 7.19 & 1.25 & 11.80  & 668.2043 & 0.2773 & 0.1318 \\
22 & 1.5 & 1.25 & 0.5 & 22.93 & 55.14 & 7.69 & 1.25 & 23.58  & 673.4684 & 0.3232 & 0.1215 \\
23 & 1.5 & 1.25 & 1   & 22.23 & 53.40 & 8.19 & 1.25 & 31.16  & 673.8875 & 0.3400 & 0.1158 \\
24 & 1.5 & 1.75 & 0   & 23.84 & 57.33 & 7.19 & 1.75 & 11.83  & 668.0523 & 0.2751 & 0.1321 \\
25 & 1.5 & 1.75 & 0.5 & 22.97 & 54.41 & 7.69 & 1.75 & 23.08  & 671.2299 & 0.3157 & 0.1238 \\
26 & 1.5 & 1.75 & 1   & 22.29 & 52.77 & 8.19 & 1.75 & 30.50  & 670.3263 & 0.3306 & 0.1192 \\
\bottomrule
\end{tabular}
\end{adjustbox}
\label{tab:houston_policy}
\end{table}



\clearpage
\section{Sample Prompt}

\begin{figure}[H]
    \centering
    \includegraphics[width=0.95\linewidth]{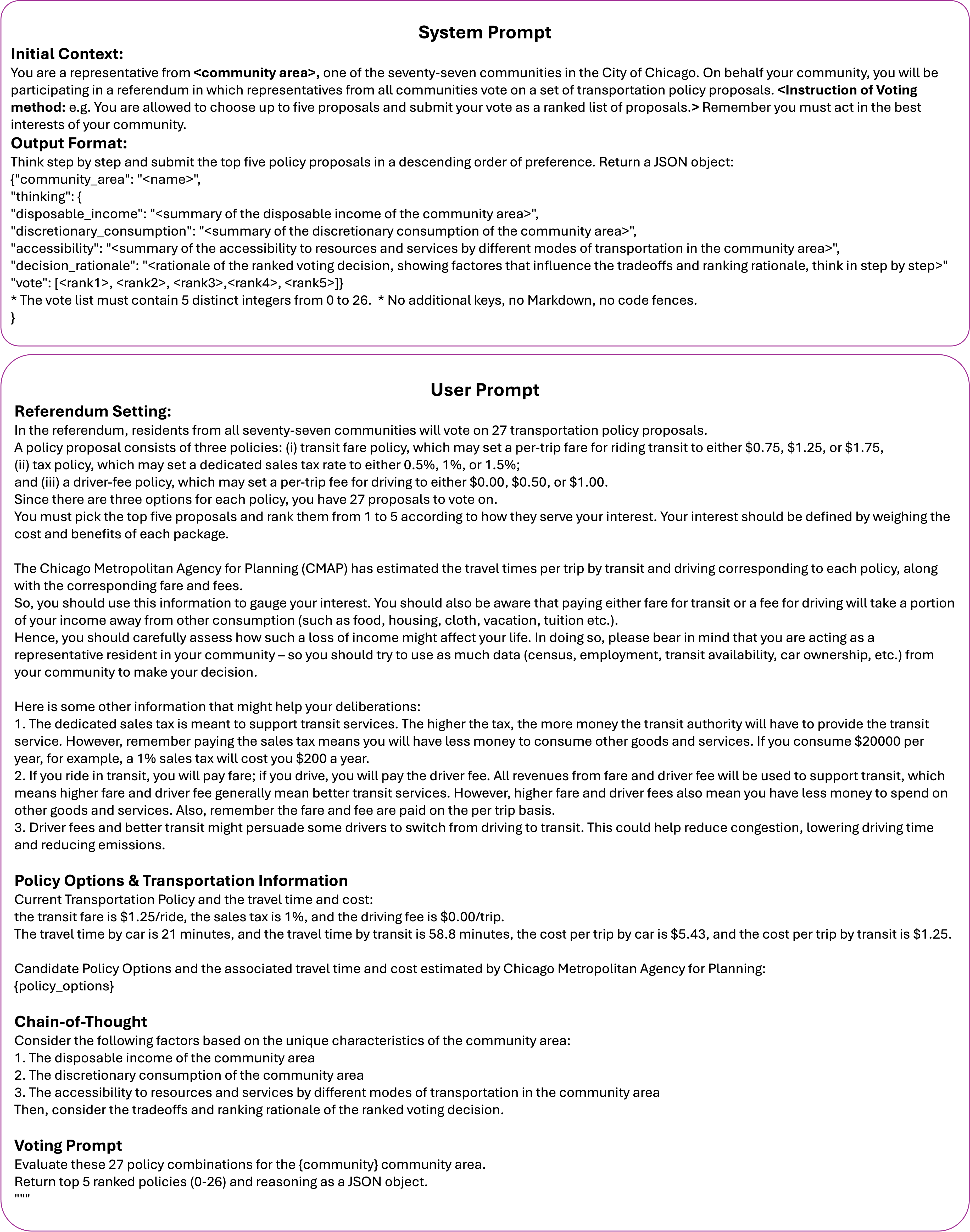}
    \caption{Example Prompt.}
    \label{fig:prompt_exp}
\end{figure}

\subsection{Sample response of All-Approval by GPT-4o}
\label{sec:sample_response_all_approve}
Community Area: Loop\\
Chain-of-Thought:
\begin{enumerate}
    \item Disposable Income: ``The Loop is a central business district with a mix of high-income professionals and lower-income service workers. Disposable income varies significantly, but there is a substantial portion of residents with higher disposable income."
    \item Discretionary Consumption: ``Residents in the Loop have a high level of discretionary consumption due to the presence of numerous dining, entertainment, and shopping options. However, the cost of living is also high, which can impact discretionary spending."
    \item Accessibility: ``The Loop has excellent accessibility to resources and services, with extensive public transit options including buses and trains. Car ownership is less common due to the availability of public transit and the high cost of parking."
    \item Decision Rationale: ``Given the Loop's strong public transit infrastructure and the presence of high-income residents, policies that enhance transit services while maintaining reasonable costs are preferred. A lower transit fare and moderate driver fees can encourage transit use and reduce congestion. A moderate sales tax is acceptable if it significantly improves transit services, as many residents can absorb the cost. Therefore, policies with a \$0.75 fare and a \$0.5 driver fee, combined with a 1\% or 1.5\% sales tax, are prioritized to balance cost and service improvement."
\end{enumerate}
Vote Decision: [10, 11, 19, 20]
\noindent
\subsection{Sample response of 5-Approval by GPT-4o}

Community Area: Loop\\
Chain-of-Thought:
\begin{enumerate}
    \item Disposable Income: ``The Loop is a central business district with a mix of high-income professionals and lower-income service workers. Disposable income varies significantly, but there is a substantial portion of residents with higher disposable income."
    \item Discretionary Consumption: ``Residents in the Loop have a higher level of discretionary consumption due to the presence of numerous dining, entertainment, and shopping options. However, the cost of living is also high, which affects overall discretionary spending."
    \item Accessibility: ``The Loop has excellent accessibility to resources and services, with extensive public transit options including buses and trains. Car ownership is lower compared to other areas due to the availability of public transit and the high cost of parking."
    \item Decision Rationale: ``Given the Loop's strong public transit infrastructure and the mixed income levels, policies that enhance transit services without significantly increasing costs for lower-income residents are preferred. A lower transit fare is beneficial for encouraging public transit use, while a moderate driver fee can help reduce congestion and emissions. A moderate sales tax is acceptable if it significantly improves transit services. Therefore, policies with lower fares and moderate driver fees are prioritized, while ensuring the sales tax does not overly burden residents."
\end{enumerate}
\noindent
Vote Decision: [1, 4, 10, 13, 19].

\subsection{Sample response of ranked-choice voting by GPT-4o}
\label{sec:sample_response_rk}
Community Area: Loop\\
Chain-of-Thought:
\begin{enumerate}
    \item Disposable Income: ``The Loop is a central business district with a mix of high-income professionals and lower-income service workers. Disposable income varies significantly, but overall, the area has a relatively high average income."
    \item Discretionary Consumption: ``Residents in the Loop have a high level of discretionary consumption due to the presence of numerous dining, shopping, and entertainment options. However, the cost of living is also high, which affects disposable income."
    \item Accessibility: ``The Loop has excellent accessibility to resources and services, with extensive public transit options including buses and trains. Walking and biking are also viable options due to the dense urban environment."
    \item Decision Rationale: ``Given the high accessibility to public transit and the mixed income levels, the Loop community would benefit from policies that enhance transit services while keeping costs manageable. Lower transit fares and moderate driver fees can encourage more transit use, reducing congestion and emissions. A moderate sales tax is acceptable if it significantly improves transit services. The focus is on balancing cost with improved transit efficiency and reduced travel times."
\end{enumerate}
\noindent
Voting Decision: [11, 14, 13, 10, 19]

\end{document}

%% file: math_commands.tex
\usepackage{amsmath,amsfonts,bm}

\def\1{\bm{1}}

\def\vbeta{{\bm{\beta}}}

\def\vp{{\bm{p}}}

\def\vv{{\bm{v}}}

\DeclareMathAlphabet{\mathsfit}{\encodingdefault}{\sfdefault}{m}{sl}
\SetMathAlphabet{\mathsfit}{bold}{\encodingdefault}{\sfdefault}{bx}{n}